\documentclass[11pt]{article}
\usepackage{jheppub1}
\usepackage{amsmath,amssymb,amsfonts,graphicx,subfigure}

\newcommand{\beq}{\begin{equation}}
\newcommand{\eeq}{\end{equation}}
\def\bea{\begin{eqnarray}}
\def\eea{\end{eqnarray}}

\def\({\left(}
\def\){\right)}

\def\bea{\begin{eqnarray}}
\def\eea{\end{eqnarray}}
\def\nn{\nonumber}
\def\ba{\begin{array}}
\def\ea{\end{array}}
\def\nn{\nonumber}
\def\Tr{\text{Tr}}
\def\Sch{\text{Sch}}
\def\sgn{\text{sgn}}
\def\J{\mathcal{J}}
\def\V{\mathcal{V}}
\def\S{\mathcal{S}}

\def\O{\mathcal{O}}
\def\T{\mathcal{T}}
\def\N{\mathcal{N}}

\def\Pf{\text{Pf}}
\def\ii{\mathfrak{i}}
\def\t{\tau}

\title{Chaos-protected locality}

\author[]{Shao-Kai Jian,}

\author[]{Brian Swingle}

\affiliation[]{Department of Physics, Brandeis University, Waltham, Massachusetts 02453, USA}

\affiliation[]{Condensed Matter Theory Center and Joint Quantum Institute, Department of Physics, University of Maryland, College Park, Maryland 20742, USA}

\emailAdd{skjian@brandeis.edu, bswingle@umd.edu}

\abstract{Microscopic speed limits that constrain the motion of matter, energy, and information abound in physics, from the ``ultimate'' speed limit set by light to Lieb-Robinson speed limits in quantum spin systems. In addition to these state-independent speed limits, systems can also be governed by emergent state-dependent speed limits indicating slow dynamics arising, for example, from slow low-energy quasiparticles. Here we describe a different kind of speed limit: a situation where complex information/entanglement spreads rapidly, in a fashion inconsistent with any speed limit, but where simple signals continue to obey an approximate speed limit. If we take the point of view that the motion of simple signals defines the local spacetime geometry of the universe, then the effects we describe show that spacetime locality can be compatible with a high degree of non-local interactions provided these are sufficiently chaotic. With this perspective, we sharpen a puzzle about black holes recently raised by Shor and propose a schematic resolution.}


\begin{document}
\maketitle

\newpage
\parskip=10pt

\section{Introduction}

A basic property of the spacetime geometry of the universe is locality: matter, energy, and information cannot be conveyed from one local point to another distant point instantaneously. 
Motion from an emitter to a receiver requires a non-zero propagation time equal to the distance between the two divided by the speed light. 
Suppose, however, that there were interactions in nature that violated this locality rule by directly coupling distant points. 
Would such interactions necessarily spoil the physics of spacetime locality? 
In this work, we show that the answer to this question is no. 
This is achieved by exhibiting a model in which there are totally non-local interactions, yet simple local signals approximately obey a speed-of-light-like speed limit.

The standard intuition is that such non-local interactions would be easily detectable since they would permit quantum information to be moved essentially instantaneously. 
This intuition amounts to an implicit model of the non-local interactions as simple short cuts, wormholes of a sort, through which, say, photons can easily propagate. Upon further thought, however, the situation is not so clear: would such non-local connections be generically detectable by local observers?

Intuitively, if simple signals were scrambled upon passing through a would-be shortcut, then it might be very hard to detect that information was being spread non-locally. 
In other words, if the local structure of spacetime were~\textit{defined} in terms of the propagation of simple signals (created, manipulated, and detected using local equipment built from relatively small sets of degrees of freedom), then it might be the case that non-local connections would be ineffective at propagating such signals in a detectable form. 
Hence, simple signals could approximately obey the causal structure of a local spacetime, while more complex (and harder to detect) forms of information and entanglement could spread rapidly in a fashion inconsistent with local causality.

We show that it is indeed possible for non-local couplings to respect the locality structure of simple signals by exhibiting a model with the desired physics. 
More precisely, if the local structure of spacetime is defined using the propagation of simple signals\footnote{For example, we may define simple signals as those that can be created and detected using only a few degrees of freedom at a time (or at least a number not growing with system size).}, then this local structure is not strongly modified by the non-local couplings. 
We further argue that the physics exhibited by this simple model should be generic across a broader class of models. 
The key physical ingredient in the construction is quantum chaos, hence we term it chaos-protected locality.

Our considerations in this work were motivated by certain puzzles in the quantum physics of black holes, but the physics we describe is not restricted to that setting. 
In the black hole context, we will discuss a seeming conflict, recently highlighted by Shor~\cite{Shor:2018scrambling}, between the local structure of a black hole spacetime and the rate of entanglement generation by the black hole. 
We argue that the physical effects described here provide a skeleton for a resolution of that puzzle by showing that rapid non-local entanglement growth can coexist with speed limits for simple signals.

The rest of the paper is organized as follows. We first give an overview of chaos-protected locality and some models that realize it in the next subsection. Then in Section~\ref{sec:model} we fully define a precise model, which is a variant of the Sachdev-Ye-Kitaev (SYK) model~\cite{Kitaev:2015simple, Maldacena:2016remarks, Sachdev:1993gapless}. We show that simple signals are carried by quasiparticles which travel at a speed limited by microscopic parameters, so that the time for a simple signal to travel through two locations is proportional to the distance. Conversely, non-local signals that cannot be detected by simple equipment, for instance those that are characterized by out-of-time order correlations (OTOCs)~\cite{Larkin:1969quasiclassical, Shenker:2013black, Kitaev:2015simple}, can spread in a much faster manner. In particular, the time for OTOCs to grow significantly between any two locations are given not by the distance but by the logarithm of the system size.

It was recently shown that the OTOC and the entanglement between parts of the system are in general characterized by different time scales~\cite{Shor:2018scrambling, Harrow:2019a}. Nevertheless, we show in Section~\ref{sec:quench} that our model is a fast scrambler~\cite{Lashkari:2011towards} in the sense that it takes time that scales logarithmically in the system size to establish large entanglement between two unentangled halves. This time scale is the same as the time scale for OTOC to be sizable. The technical calculation considers the time evolution of the second R\'enyi entropy from an untangled initial state and shows that the entanglement increases at a rate proportional to the size of subsystem, in stark contrast to a quench in a local one-dimensional theory for which the rate of entanglement growth is a constant independent of subsystem size~\cite{Calabrese:2005evolution}.

In Section~\ref{sec:BH} we review Shor's cell model description of black hole dynamics and the consequent puzzle raised by it. Shor's work distinguishes two notions of information scrambling, which was later justified more rigorously in a random quantum circuit model defined on graphs with a bottleneck~\cite{Harrow:2019a}. Strong scrambling, defined as reaching a nearly maximally entangled state starting from two weakly-entangled macroscopic parts of a system, is seemingly constrained by the causal structure of the black hole spacetime to occur too slowly relative to expectations from quantum black hole physics. We sharpen this puzzle by generalizing the cell model to a Schwarzschild-AdS black hole and referencing precise calculations of the entanglement dynamics obtained from AdS/CFT duality. Then we discuss a possible resolution of the tension based on our explicit model: the degrees of freedom at the stretched horizon are essentially non-local so that the horizon is a fast scrambler~\cite{Lashkari:2011towards} even in the sense of strong scrambling while the causal structure can be maintained from the viewpoint of an outside observer with access only to simple probes.

\subsection{Detailed overview}

\begin{figure}
\begin{center}
\includegraphics[width=.6\textwidth]{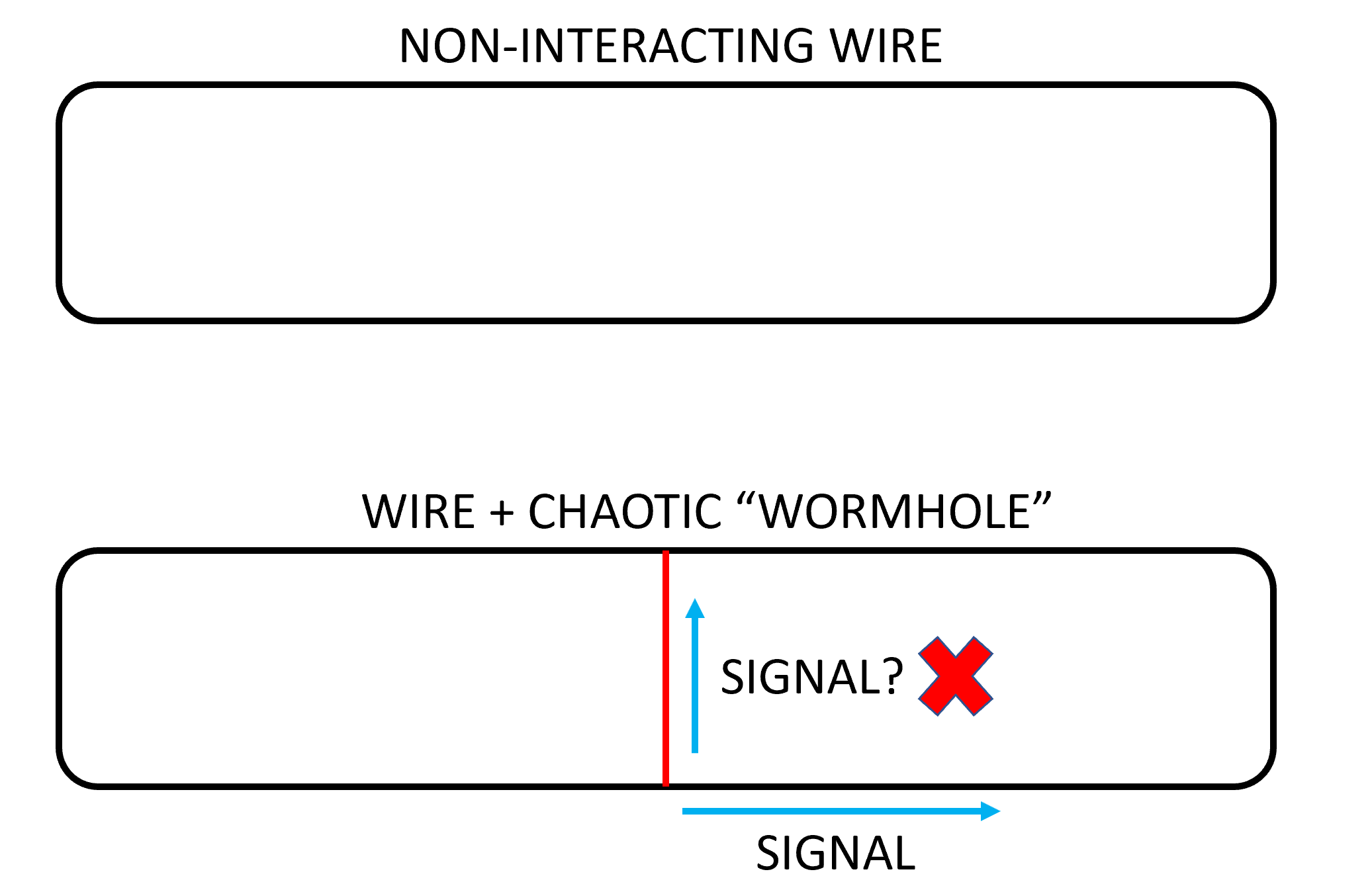}
\caption{A simplified version of chaos-protected locality. In the upper diagram, energy and information can only propagate from bottom to top the long way around. In the lower diagram, there are now in principle two routes that can be taken from bottom to top, but if the short-cut is chaotic, then it will not transmit any simple signal. Hence, if the original simple wire does host long-lived excitations, then the time needed to locally signal from bottom to top is still set by the length of the simple wire, despite the short-cut.}
\label{fig:wire-shortcut}
\end{center}
\end{figure}

A toy model of the effect is shown in Figure~\ref{fig:wire-shortcut}. 
The upper diagram depicts a simple wire which has a geometrically local time evolution with some speed limit for quantum information propagation. 
In particular, the time it takes to send a signal from the middle of the bottom of the wire to the middle of the top of the wire is long if the wire is long. 
The lower diagram depicts a modified situation in which a shortcut is added that connects the top and bottom of the simple wire. 
This shortcut wire is unlike the simple wire because it is strongly interacting and chaotic. 
In particular, while energy may slowly diffuse along the chaotic wire, no locally excited signal is able to propagate for any significant distance.\footnote{To guarantee that even sound modes are strongly damped, one could add some static disorder potential to the chaotic wire.}
This inability to propagate simple signals is a consequence of thermalization, since the chaotic system effectively forgets its initial condition, including the would-be signal, as far as few-body observables are concerned. 
More properly, the information in the signal is scrambled up into complex and essentially unobservable many-body degrees of freedom.

If the chaotic wire is significantly shorter than the simple wire, then there is a danger of observing a violation of locality in the simple wire. 
However, if the chaotic wire is significantly longer than some attenuation length scale, then no simple signal will be able to propagate from one end of the chaotic wire to the other. 
More precisely, the mutual information between local degrees of freedom at either end of the chaotic wire will be nearly zero until information is able to propagate the long way around through the simple wire. 
Hence, locality, as measured by the time it takes to propagate simple signals, would remain approximately intact. 
Of course, the chaotic wire is spreading entanglement in a fashion inconsistent with the local structure of the simple wire, but this spreading is hard to detect with ordinary measurements. 
Roughly speaking, the chaotic wire just looks like a thermal bath which absorbs and emits energy into the simple wire.

\begin{figure}
\begin{center}
\includegraphics[width=.6\textwidth]{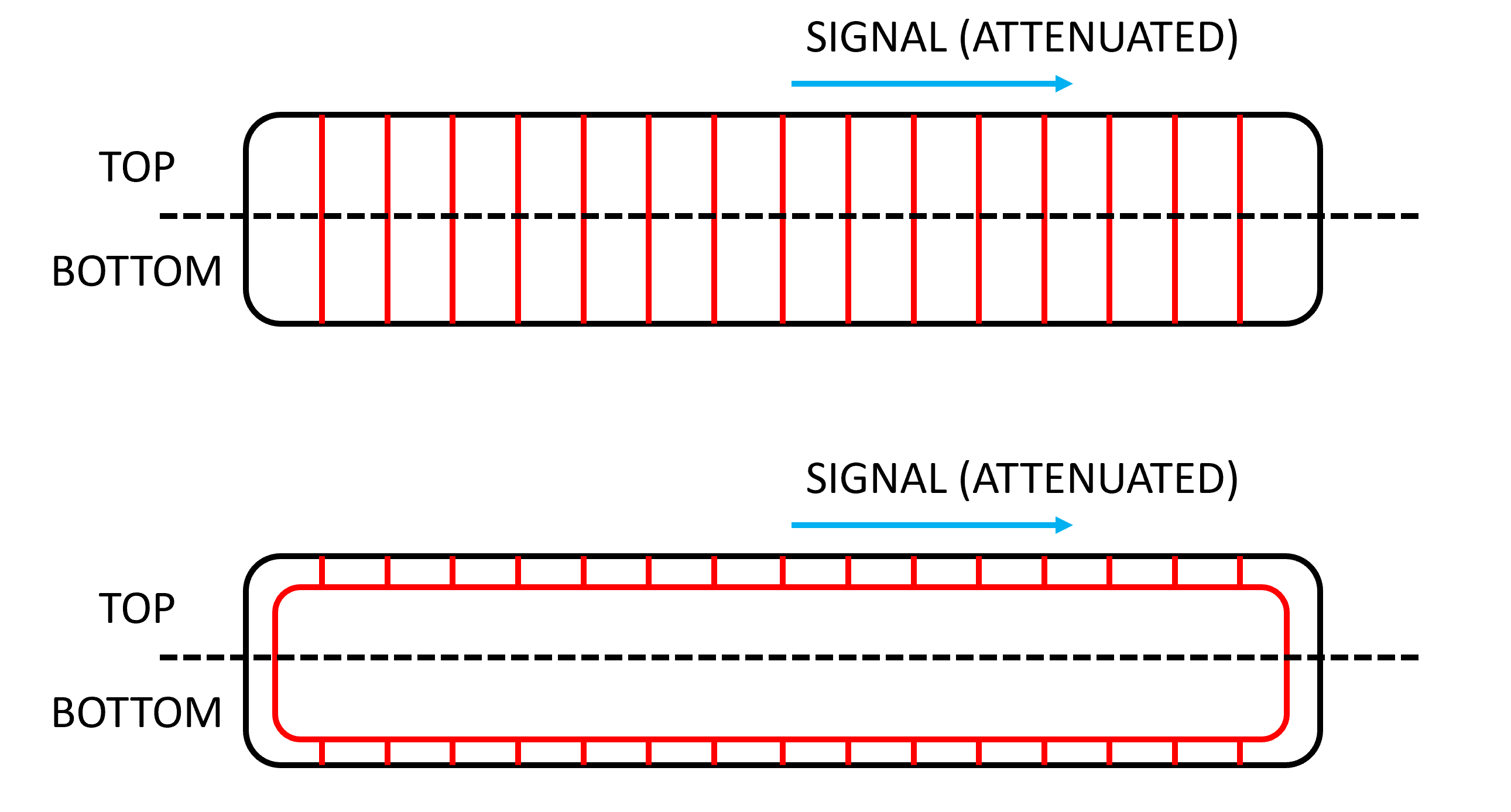}
\caption{Entanglement growth in two different situations. In the upper diagram, entanglement between the top and the bottom will grow rapidly, at a rate proportional to the length of the wire. However, the chaotic wires will still not be able to convey any simple signal. In the lower diagram, attenuation of simple signals will continue to take place as in situation in the upper diagram, but now the system obeys a microscopic speed limit and entanglement between the top and the bottom will grow much more slowly, at a constant rate independent of wire length.}
\label{fig:wire-fast-entanglement}
\end{center}
\end{figure}

Generalizing this setup, consider the two possible situations shown in Figure~\ref{fig:wire-fast-entanglement}. 
In the upper panel, there are connections between the simple wire and many chaotic wires. 
These connections violate locality as defined by the simple wire's Hamiltonian. 
However, at least if the coupling between the two wire types is weak or irrelevant in the renormalization group sense at low energies, signals will continue to propagate locally in the simple wire, up to a small attenuation induced by the coupling to the chaotic wires. By contrast, entanglement between the top and bottom degrees of freedom (the entanglement bipartition is indicated by the dashed line in Figure~\ref{fig:wire-fast-entanglement}) will grow rapidly at a rate proportional to the length of the simple wire. 
Without the locality violating connections, entanglement in the simple wire would have grown at a rate proportional to the size of the cut between top and bottom, which would be independent of the length. 
By contrast, in the scenario shown in the lower panel of Figure~\ref{fig:wire-fast-entanglement}, signals in the simple wire will still be attenuated by the coupling to the chaotic wires, but now the chaotic wires actually obey the local structure of the simple wire and the entanglement growth between the top and bottom will be much slower.

\begin{figure}
\begin{center}
\includegraphics[width=.5\textwidth]{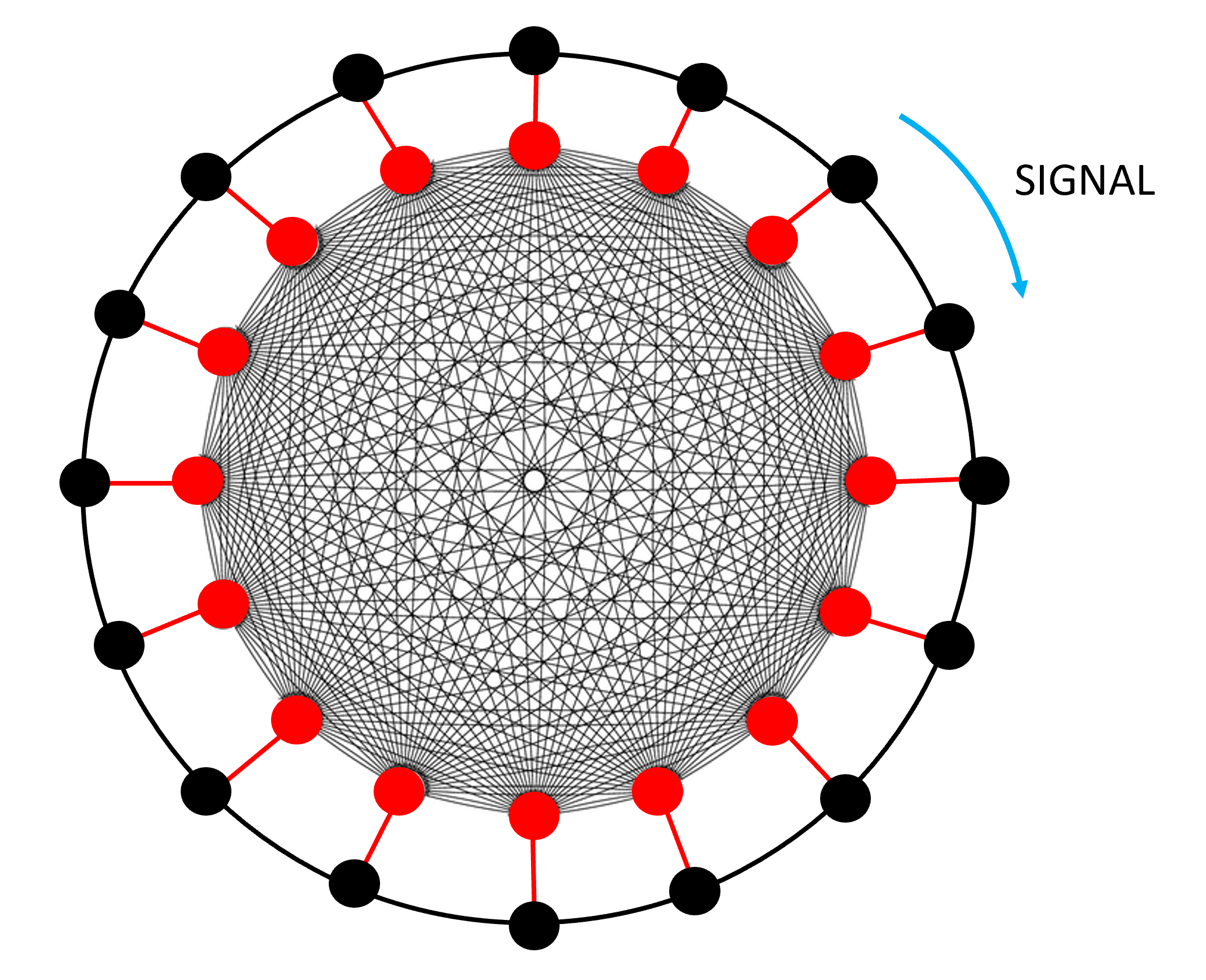}
\caption{All-to-all coupled model. The outer black ring plays the role of a simple wire. In the model discussed below, it will correspond to non-interacting Majorana fermions which we also call local Majorana. The inner red ring plays the role of the chaotic non-local shortcut. In the model below, it will correspond to an all-to-all interacting SYK model which we call chaotic Majorana. As in the wire models, the key physical features of this setup are that simple signals continue to obey a local speed limit while complex entanglement will spread rapidly.}
\label{fig:wire-complete}
\end{center}
\end{figure}

The wire model gives good intuition, but a more complex setup is desired in which the analog of the chaotic wires is played by a set of degrees of freedom with chaotic all-to-all interactions. 
The basic physical situation is illustrated in Figure~\ref{fig:wire-complete}. 
There the outer ring of black sites is the analog of the simple wire in which signals can propagate freely. 
The inner red sites are all-to-all coupled as indicated. 
This model can be augmented by adding more simple degrees of freedom outside the outer ring, so that one has a higher-dimensional space of simple degrees of freedom coupled at a boundary of that space to some non-locally interacting chaotic modes. 
Note that a geometrical extension of the basic picture sketched in Figure~\ref{fig:wire-complete} where one adds additional simple degrees of freedom outside is motivated by the aforementioned black hole puzzle~\cite{Shor:2018scrambling}.

It will be shown that for a certain choice of Hamiltonian, the situation in Figure~\ref{fig:wire-complete} has the following properties. 
First, the local structure of the outer ring is preserved up to corrections that are suppressed by a power of the system size. 
Hence, for a large system, the corrections to the local structure are very small. 
This is so for a broad class of states, including thermal equilibrium states, even though the microscopic Hamiltonian has no ballistic Lieb-Robinson bound. 
Second, the chaotic non-local couplings do permit the overall system to scramble and spread entanglement faster than would be possible in any local Hamiltonian.\footnote{Here and throughout, we use locality to mean geometric locality. Interactions involving a small number of degrees of freedom at a time are termed few-body interactions (or $k$-body interactions if $k$ degrees of freedom are involved).} 
In essence, the inner chaotic ring acts like a thermal bath that exchanges energy with the simple wire, but does not otherwise appear to have any special non-local structure.

To describe the physics in more detail, we first introduce some notation. 
Denote the local system by $L$ and the chaotic system by $C$ and let the coupling between the two be $V$. 
For concreteness, in this paper, $L$ will be taken to be a chain of non-interacting Majorana fermions and $C$ will be taken to be an instance of the SYK model~\cite{Kitaev:2015simple, Maldacena:2016remarks, Sachdev:1993gapless}, with the details of the model to be specified in the next section. 
The main idea of the construction is as follows. 
The local system $L$ is a model of non-interacting particles. 
Being non-interacting, wavepackets of these particles are able to travel unencumbered for great distances. 
However, the particles cannot travel faster than a certain fixed velocity, which defines a local structure for the system $L$.
By contrast, the chaotic system $C$ has no notion of locality or propagation in space and does not host any weakly interacting excitations.
It is a quantum chaotic system that very quickly effectively loses memory of its initial state and approaches an effective thermal equilibrium state appropriate to its overall energy.

If the systems are coupled, then the microscopic locality of $L$ is immediately lost. 
However, as argued above, the system is expected to maintain some effective sense of locality. 
At weak coupling, three important physical effects will be induced. 
First, signals propagating in $L$ will be damped by the coupling to $C$. 
The amplitude of such signals will decay exponentially with time at a rate $\gamma$ scaling like $V^2$ at small $V$ (from Fermi's golden rule). 
Second, $C$ will induce thermal noise in $L$ as if $L$ were coupled to a heat bath of the appropriate temperature. 
Hence, even if $L$ is initially in its vacuum state, it will come to local equilibrium with $C$ after a time of order $1/\gamma \sim 1/V^2$ (this is the local equilibration time).
Third, there is some amplitude for signals in $L$ to propagate non-locally via $C$ to some distant region of $L$. 
This process is suppressed by $V^2$ at weak coupling, but more importantly, it is suppressed by a power of $N$, the number of degrees of freedom in $C$\footnote{We take both numbers of local and chaotic Majorana to be $N$ in the discussion for simplicity.}.

On the other hand, the growth of operators can be extremely rapid in the combined $LC$ system. 
For example, if out-of-time-order correlators (OTOCs) in $C$ exhibit exponentially fast growth with rate $\lambda$, then we expect OTOCs in $L$ will also grow exponentially at the same rate, only with the prefactor of the exponential reduced by a factor of $V^4$. 
This is shown by noting that $L$ operators can be converted to $C$ operators by commuting them with the $LC$ coupling which is proportional to $V$. 
Since the OTOC is a four-point function, four powers of $H_{LC}$ are needed corresponding to four powers of $V$. 
The timescale to fully scramble the $C$ system would then be $\lambda^{-1} \log M$ and the timescale to fully scramble $LC$ will be only slightly longer, of order $\lambda^{-1} \log \frac{N}{V^4}$. 
Similarly, the timescale to nearly maximally entangle, say within a few bits of maximal, the top half of the system shown in Figure~\ref{fig:wire-complete} with the bottom half will be of order $\lambda^{-1} \log N$ for $C$ alone and not much longer for the combined $LC$ system.\footnote{In principle there could be another timescale distinct from $\lambda^{-1}$ here, but for simplicity of presentation they are assumed to be of the same order. Also, at non-infinite temperature one should replace nearly maximal entanglement with the amount of entanglement appropriate to a pure thermal-like state.} 
This is much shorter than the $O(N)$ time it would take $L$ to reach nearly maximal entanglement without the aid of $C$.

These timescales are estimates at weak coupling, but we expect the system size dependence to remain the same even at strong coupling. 
In this case, however, the attenuation of simple signals will be strong and local communication will anyway be difficult over any appreciable distance. 
At weak coupling, we will explicitly show the large $N$ suppression of non-local signalling below. 
The entanglement growth claims will also be explicitly demonstrated here by studying the time evolution of the second R\'enyi entropy of a global quench from an unentangled initial state. 
With the help of the $LC$ model, we show the entanglement growth rate is proportional to the number of degrees of freedom, indicating a much faster scrambling ability for non-local signals. 

To summarize, there is a physical setup where simple signals in $L$ obey the ordinary rules of local causality, up to a small attenuation induced by $C$. 
Such signals do have a small amplitude to propagate non-locally, but this is hard to distinguish from the thermal noise being emitted by $C$ into $L$. 
On the other hand, operator growth and entanglement growth are extremely rapid, but without multiple copies of the system or the ability to control the flow of time, observers will have difficulty detecting these features of the physics. 
In other words, simple observers with access to small parts of $L$ cannot detect rapid entanglement growth; only super-observers who have access to and fine control over the complete system can hope to see such non-local effects. The basic outline of the physics should remain true even if the coupling $V$ is not weak, but the situation is harder to analyze.

As indicated above, the model is intended to capture the near horizon quantum dynamics of a black hole. 
The $L$ system is analogous to photons or some other weakly coupled degrees of freedom outside the horizon of the black hole. 
It is a stand in for the Hawking radiation of the black hole. 
The $C$ system corresponds to the stretched horizon, which is highly chaotic and sensitive to the physics of quantum gravity. 
Thus, we propose that high energy quantum gravity degrees of freedom could strongly violate spacetime locality without any significant indication of this violation in the Hawking radiation. 
Moreover, we will argue that this sort of non-locality at the stretched horizon is necessary, at least for black holes similar to those in Anti de Sitter spacetime, connecting to a recent discussion due to Shor. 
More generally, it has long been known in simple models of quantum gravity that non-locally interacting ``matrix'' degrees of freedom somehow generate local dynamics, but the precise mechanisms remain mysterious. 
We consider the effects discussed here as a small step towards unraveling this mystery by showing that locality can be surprisingly robust to non-local perturbations.

\section{The $LC$ model} \label{sec:model}

We consider the following variant of the SYK model~\cite{Kitaev:2015simple, Maldacena:2016remarks, Sachdev:1993gapless}, which we refer to as $LC$ model, consisting of $N$ local Majorana and $M$ chaotic Majorana,
\bea
	H &=&  H_L + H_C + H_{LC}, \\
\label{eq:chain}	H_L &=& - \ii w \sum_r \psi_r \psi_{r+1}, \\
	H_C &=& \ii^{q/2}  \sum_{i_1<...<i_q} J_{i_1,...,i_q} \chi_{i_1}...\chi_{i_q}, \\
	H_{LC} &=& \ii^{(p+1)/2} \sum_{r, i_1<...<i_p}V_{r,i_1,...,i_p} \psi_r \chi_{i_1} ... \chi_{i_p},
\eea
where $\psi_r$, $r=1,...,N$ denotes the local Majorana at site $r$ and we implement periodic boundary condition $\psi_{r+N}= \psi_r$, and $\chi_i$, $i=1,...M$ denotes the chaotic Majorana. 
$q$ is an even integer representing the $q$-body all-to-all interaction of the chaotic Majorana, and $p$ is an old integer representing the $(p+1)$-body all-to-all interaction between $p$ chaotic Majorana and a local Majorana. 
$w$ is the hopping amplitude for the local Majorana. 
$J_{i_1,...,i_q}$ and $V_{r,i_1,...,i_p}$ are Gaussian random variables with mean zero and variance given by
\bea
\label{eq:J2}	\overline{J_{i_1,...,i_q}J_{i'_1,...,i'_q}} &=& \delta_{i_1,i'_1}...\delta_{i_q,i'_q} \frac{(q-1)! J^2}{M^{q-1}} = \delta_{i_1,i'_1}...\delta_{i_q,i'_q} \frac{(q-1)! 2^{q-1} \J^2}{q M^{q-1}}, \\
\label{eq:V2}	\overline{V_{r,i_1,...,i_p}V_{r',i'_1,...,i'_p}} &=& \delta_{r,r'}\delta_{i_1,i'_1}...\delta_{i_p,i'_p} \frac{(p-1)! V^2}{M^{p}} = \delta_{r,r'} \delta_{i_1,i'_1}...\delta_{i_p,i'_p} \frac{(p-1)! 2^{p-1} \V^2}{p M^{p}}.
\eea
The above equations also define the effective interaction strength $J$ and $V$. 
The scaled constants $\J$ and $\V$ are defined for the purpose of the large $q$ and $p$ limits, as discussed below. When $N=M$ and $p=1$, the $LC$ model represents an example of Fig.~\ref{fig:wire-complete}.

We first investigate the equilibrium properties of the model. The imaginary time path integral that represents the thermal partition function is controlled by the action
\bea
	I &=& I_C + I_L + I_{LC} \\
	- \frac{I_C}M &=& \log \Pf(\partial - \Sigma_\chi) + \int d\t_1 d\t_2\Big[ - \frac12 G_\chi(\t_1,\t_2) \Sigma_\chi(\t_1,\t_2) + \frac{J^2}{2q} G_\chi(\t_1,\t_2)^q \Big] \\
	- I_L &=&  - \frac12 \int d\t \sum_{r,r'}\psi_r(\t)(  \partial_\t \delta_{r,r'}- \ii 2 w h_{r,r'}) \psi_{r'}(\t) \\
	- I_{LC} &=&  \frac{V^2}{2p} \int d\t_1 d\t_2 \sum_r \psi_r(\t_1) G_\chi(\t_1,\t_2)^p \psi_r(\t_2),
\eea
where $h_{rr'} = \frac12 (\delta_{r,r'+1} + \delta_{r,r'-1})$ is the hopping matrix of the local Majorana.
$G_\chi(\t_1, \t_2) = \frac1{M} \sum_i \chi_i(\t_1) \chi_i(\t_2)$ and $\Sigma_\chi(\t_1, \t_2)$ are the propagator and the self-energy of the chaotic Majorana. 
Integrating out the local Majorana fermion, the equation of motion for these bilocal fields reads
\bea
\label{eq:eom1}     G_\chi^{-1} &=& \partial -\Sigma_\chi, \\
\label{eq:eom2}	 \Sigma_\chi(\t_1, \t_2) &=& J^2 G_\chi(\t_1, \t_2)^{q-1}  \nn \\
	&& + \frac{V^2}{M} \Tr \left[ \left( [\partial_{\t_1} - \ii 2 w h]\delta(\t_1-\t_2) - \frac{V^2}p G_\chi(\t_1,\t_2)^p  \right)^{-1} \right] G_\chi(\t_1,\t_2)^{p-1}. 
\eea
The trace in the second equation is over the local Majorana indices $r$.

In the following sections, we will first consider the simple and non-local signals for the local Majorana in the limit $M/N \gg 1$, where the backreaction from the local Majorana to the chaotic Majorana is negligible. We will then discuss the effect of backreaction.

\subsection{Simple signal: two-point correlation function} \label{subsec:simple_signal}

Without coupling to the chaotic Majorana, the local Majorana is a free model that can be solved by Fourier transform,
\bea
	\psi_k = \frac1{\sqrt N} \sum_k \psi_r e^{\ii k r}, \quad k = -\pi + \frac{2\pi n}{N}, \quad n= 1,..,N .
\eea
Various propagators of the free Majorana model are given in Appendix~\ref{append:free}. The thermal propagator is given by
\bea
	G_{\psi,0}(\omega,k) = \frac1{-\ii \omega + \varepsilon_k}, \quad \varepsilon_k = 2w \sin k.
\eea
It is not hard to see from above propagator that the fastest velocity that a local Majorana fermion can move is given by $v \equiv 2 w$. Indeed, if we look at the low-energy limit, $w \beta \gg 1$, the retarded Green's function is
\bea\label{eq:retard}
	\theta(t) \langle \{\psi_{r}(t), \psi_{r'}(0) \} \rangle = - \frac{\ii}{\pi v} \sum_{s=\pm 1} \frac{\pi}{\beta \sinh \frac{\pi(t- s \frac{r-r'}{v})}{\beta}},
\eea
which means that for a local Majorana to propagate from position $r'$ to $r$, the time it costs is $ t = \frac{|r-r'|}v$, because for any time shorter than that, the signal is exponentially suppressed.

Now let us discuss the effect of the coupling to the chaotic Majorana. In the limit $M/N \gg 1$ such that the backreaction can be neglected, the equation of motion~(\ref{eq:eom1}, \ref{eq:eom2}) reduces to the SYK model, and the solution in the conformal limit is~\cite{Maldacena:2016remarks}
\bea
	G_\chi(\tau) = \sgn(\t) \frac{b}{|\tau|^{2/q}}, \quad b^q = (\pi J^2)^{-1} \Big(\frac12 - \frac1q \Big) \tan \frac\pi{q}. 
\eea
The coupling to local Majorana is $p$ powers of the chaotic Majorana correlator,
\bea
	G_\chi^p(\tau) &=& \sgn(\t) \frac{b^p}{|\tau|^{2p/q}}, \\
	G_\chi^p(\omega) &=& \tilde b\, \ii \,\sgn(\omega) |\omega|^{2p/q-1},  \quad \tilde b = b^p 2 \cos\left(\frac{p\pi}q \right) \Gamma\left(1-\frac{2p}q \right),
\eea
where the second equation is obtained by Fourier transform. 

This leads to a bath generated self-energy for the local Majorana $\psi$,
\bea
	G_\psi^{-1}(\omega, k) &=& -\ii \omega - \frac{V^2 }{p}  \tilde b \, \ii \, \sgn(\omega) |\omega|^{2p/q-1} + \varepsilon_k.
\eea
It is not hard to show that the quasiparticle life-time for an excitation with energy $\varepsilon$ is given by 
\bea
	\tau^{-1}_{qp} = \frac{2V^2}p b^q \cos \frac{p\pi}q \sin \frac{p\pi}q \Gamma \left(1- \frac{2p}q \right) |\varepsilon|^{2p/q-1}.
\eea
In particular, when $p>q$, the quasiparticle is well defined at low energies, $\varepsilon \rightarrow 0$, and it propagates at the maximal velocity $v = 2 w$. 
So we will focus on $p>q$ case, where the chaotic Majorana leads to an irrelevant perturbation to the simple signal propagator. Surprisingly, we will next show that while this simple locality structure is approximately preserved, OTOCs can spread much faster, i.e., at the maximal rate.

\begin{figure}
\centering
\subfigure[]{\label{fig:2point}
	\includegraphics[width=0.3\textwidth]{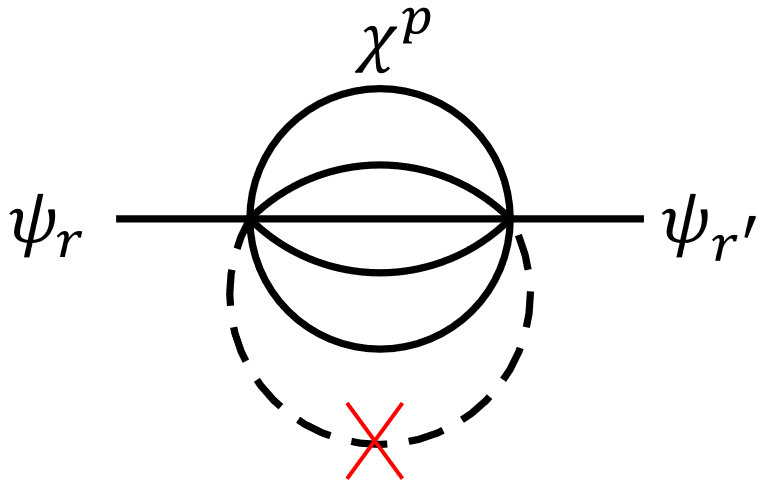}} \qquad 
\subfigure[]{\label{fig:variance}
	\includegraphics[width=0.3\textwidth]{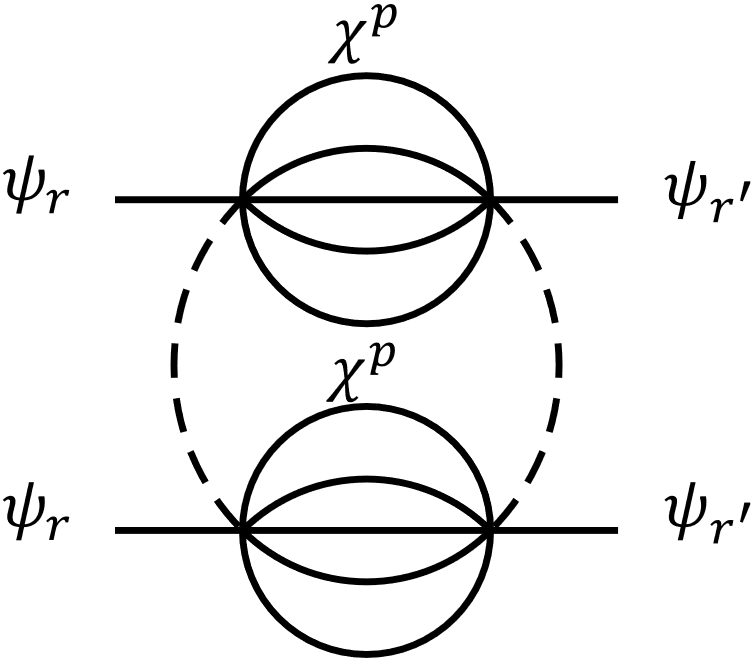}}
	\caption{(a) Under disorder average, the propagator between two local Majorana at $r \ne r'$ directly through the chaotic Majorana is zero. The dashed line represents disorder average over random variable $V_{r,j_1,...,j_p}$ (b) The variation of the propagator in (a).}
\end{figure}

Considering the disorder-averaged simple signal, we have shown that the effect of the chaotic Majorana is to produce dissipation and generate a finite life-time for the quasiparticle that is used to transmit the simple signal. One may wonder if, for a fixed disorder configuration, the signal can propagate faster. Next we argue that the fluctuation-induced signal transmission is suppressed in the large $M$ limit. For instance, Fig.~\ref{fig:2point} shows a shortcut where a simple signal can travel between two sites $r$ and $r'$ with the help of non-local couplings to the chaotic Majoranas. Although on average such apropagator between two different local Majorana $r\ne r'$ through the chaotic Majorana is zero, if we consider a fixed disorder configuration, there are still finite correlations due to the disorder fluctuation as shown in Fig.~\ref{fig:variance}. Nevertheless, one can estimate the magnitude of such a process. Each internal line of the chaotic Majorana leads to a factor of $M$ because of the summation, and the dashed line that represents the disorder average leads to a factor of $M^{-p}$ from~(\ref{eq:V2}). Moreover, the disorder average also kills $p$ out of the $2p$ summations so that there are only $p$ independent summations. Thus, the standard deviation of such a process is suppressed by $M^{-p/2}$, i.e., no simple two-point signal can travel through the shortcut if $M$ is large enough. Compared to this fluctuation-induced correction, a larger, $1/M$ correction can actually come from the reparametrization mode at low energies. 
Nevertheless, as we discuss in the next section it is not only suppressed by $1/M$ but also irrelevant at low energies.

\subsection{Non-local information: out-of-time order correlation function} \label{subsec:non-local_signal}

A useful characterization of the scrambling of information, especially non-local information, is given by the norm square of (anti)commutator~\cite{Kitaev:2015simple,Larkin:1969quasiclassical,Shenker:2013black}.
For instance, non-local information propagating originating from position $r$ and being detected at $r'$ is seen from sizable values of the following quantities,
\bea
	\langle | \{\psi_r(t), \psi_{r'}(0) \}|^2 \rangle &=& \langle \psi_{r}(t) \psi_{r'}(0) \psi_{r}(t) \psi_{r'}(0) \rangle +  \langle \psi_{r'}(0) \psi_{r}(t) \psi_{r'}(0) \psi_{r}(t) \rangle \nn\\
	&& + \langle  \psi_{r}(t) \psi_{r'}(0) \psi_{r'}(0) \psi_{r}(t) \rangle + \langle \psi_{r'}(0) \psi_{r}(t) \psi_{r}(t) \psi_{r'}(0) \rangle,
\eea
where the first line represents the OTOC which reflects scrambling, and the rest, the time ordered correlators, are decaying functions of time. More precisely, we are interested in the four-point function of local Majorana at position $r$ and $r'$,
\bea
	F_{r,r'}(\t_1,\t_2,\t_3,\t_4) = \langle \T \psi_r(\t_1) \psi_r(\t_2)  \psi_{r'}(\t_3) \psi_{r'}(\t_4)\rangle.
\eea
where $\T$ means the imaginary time ordering. 
We will omit the imaginary time ordering in the following for notational simplicity.
Its real part is responsible for the increase of $\langle | \{\psi_r(t), \psi_{r'}(0) \}|^2 \rangle \ni  -2 \Re[F_{r,r'}(\beta + \ii t, \beta/2 + \ii t, 3\beta/4 , \beta/4)]$. 
The minus sign comes from the exchange of Majorana operators. 

Without the coupling to the chaotic Majorana, the OTOC reads
\bea
	&& F^{(0)}_{r,r'}(\t_1, \t_2, \t_3, \t_4)  \approx \frac1{(v \beta)^2}  \left(1 - \sum_{s = \pm 1}  \frac1{\cosh^2 \frac{2\pi}\beta (t- s \frac{r-r'}v)} \right). 
\eea
We leave the derivation in Appendix~\ref{append:free}. It is easy to see that the information is carried by quasiparticles traveling at speed $v$. 

\begin{figure}
\centering
\includegraphics[width=0.3\textwidth]{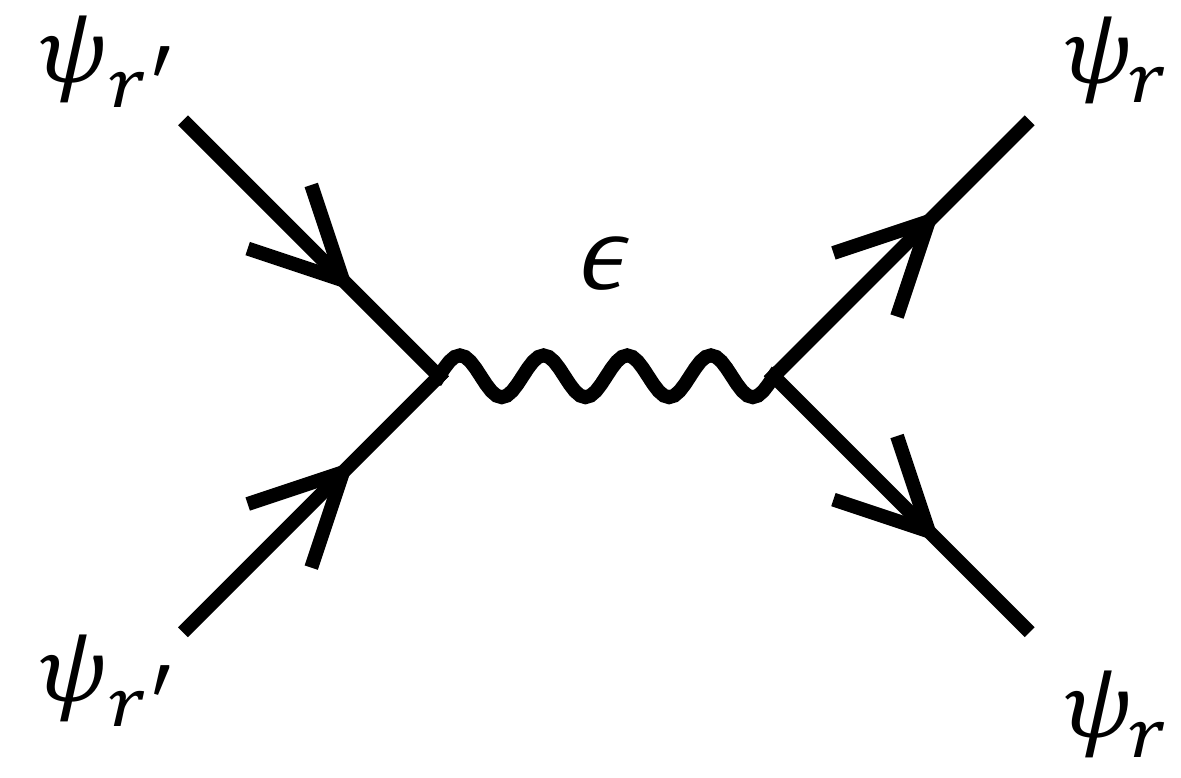}
\caption{\label{fig:otoc} The process of scrambling that conducts by the reparametrization mode that is denoted by $\epsilon$.}
\end{figure}

The coupling to chaotic Majorana will lead to the process corresponding to Fig.~\ref{fig:otoc}. 
The information at position $r'$ does not have to be carried by the quasiparticle, instead they can transmit to the chaotic Majorana though the coupling $I_{LC}$. 
Thanks to the fast scrambling nature of the chaotic Majorana, the time for non-local information to travel through half of the chain will scale logarithmically with the number of the chaotic Majorana that is independent of the length of the chain. 
In principle, the transmission between two Majorana can happen at any point between the initial and final position, but the shortest path would be transmit the information right at $r'$ and $r$, any other point will cost further time. 
So the dominant four-point function from Fig.~\ref{fig:otoc} is
\bea\label{eq:v4}
 F^{(1)}_{r,r'}(\t_1, \t_2, \t_3, \t_4) = \frac{V^4}{4p^2} \langle G^p_\chi(\tau_1, \tau_2) G^p_\chi(\tau_3, \tau_4) \rangle,
\eea
where the bracket denotes integrating over the chaotic Majorana fluctuations.
Note that the right-hand side does not depend on the position $r$ or $r'$ after disorder average.

The calculation then is similar to the OTOC in the original SYK model. We will sketch the calculation here. In the conformal limit, the chaotic Majorana is controlled by the Schwarzian action~\cite{Maldacena:2016remarks},
\bea
	-\frac{I_C}{M} = \frac{\alpha_S}{\J} \int d\t \Sch \left[ \tan \frac{ g(\t)}2, \t \right],
\eea
where $\alpha_S$ is a positive constant, $g(\tau)$ is an arbitrary reparametrization function of $\tau$, and  $\{ f(\tau), \tau \} = \frac{f'''(\tau)}{f'(\tau)} - \frac32 \left( \frac{f''(\tau)}{f'(\tau)}\right)^2$ is the Schwarzian derivative. 
We set $\beta=2\pi$ for simplicity and restore it back at the end of the calculation.
In terms of the infinitesimal reparametrization $g(\tau) = \tau + \epsilon(\tau)$, the quadratic action is
\bea
	\frac{I_C}{M} &=& \frac{\alpha_S}{2\J} \int d\t [ \epsilon''(\tau)^2 - \epsilon'(\tau)^2 ], \\
	D(\t) &=& \langle \epsilon(\t) \epsilon(0) \rangle = \frac{\J}{M \alpha_S} \left( -\frac{(|\t|-\pi)^2}2 + (|\t|-\pi) \sin |\t| + \frac52 \cos \t + 1 + \frac{\pi^2}6 \right).
\eea
Furthermore, the reparametrization mode couples to the conformal propagator through
\bea
	\delta G^p_\chi(\t_1,\t_2) = G^p_\chi(\tau_{12})  B(\tau_1, \tau_2), \quad B(\tau_1, \tau_2) = \frac{2p}q \left(\epsilon'(\tau_1) + \epsilon'(\tau_1)- \frac{\epsilon(\tau_1)-\epsilon(\tau_2)}{\tan \frac{\tau_{12}}2} \right).
\eea

With the help of the reparametrization mode, the OTOC gets an important contribution,
\bea
 F_{r,r'}^{(1)}(\t_1, \t_2, \t_3, \t_4) &=& \frac{V^4}{4p^2} \langle G^p_\chi(\tau_1, \tau_2) G^p_\chi(\tau_3, \tau_4) \rangle \\
 &=& \frac{V^4}{4p^2}  G^{2p}_\chi(\pi) (1 + \langle B(\tau_1 ,\tau_2 ) B(\tau_1, \tau_2) \rangle ) \\
 &=& \frac{V^4}{4 p^2} \left( \frac{\pi} {\beta \J} \right)^{2p/q}  \left(1 - \big(\frac{2p}q\big)^2 \frac{\beta \J}{M \alpha_S} \cosh \frac{2\pi t}\beta \right).
\eea
In the last step, we restored the inverse temperature $\beta$ and also take large $q$ limit as to be consistent with the coefficient from the Schwarzian action. Including this process, the OTOC of the local Majorana fermion reads
\bea
	F_{r,r'}(\t_1, \t_2, \t_3, \t_4) = \frac1{(v \beta)^2}  \Big(1 - \sum_{s = \pm 1}  \frac1{\cosh^2 \frac{2\pi}\beta (t- s \frac{r-r'}v)} \Big) - \frac{V^4}{M} \frac{\pi^{\frac{2p}q}(\beta \J)^{1-\frac{2p}q}}{\alpha_S q^2} \cosh \frac{2\pi t}\beta. \nn
\eea
If we consider $r-r'$ is half of the chain, namely, $r-r'= N/2$, the time for the non-local information to travel across the two points is
\bea
	t_* = \min\left( \frac{N}{2v} ,  \frac{\beta}{2\pi} \log \Big( \frac{\pi^{\frac{2p}q}(\beta \J)^{1-\frac{2p}q}}{4 \alpha_S q^2 (\beta v)^2} \times  \frac{M}{V^4}  \Big) \right),
\eea
where the first time is given by the channel of quasiparticles while the second one is the channel through the chaotic Majorana. 
So if $M$ is not exponentially larger than $N$, the channel through chaotic Majorana is faster.
Actually in the next section, we will see that the above conclusion hold true for $M=N$ at large $q$ limit. 
Thus, the non-local information can scramble at the fastest time $\propto \log N$ in the local Majorana model, while as we have seen in previous section, the simple signal still needs a linear time $\propto N$ to travel through half of the chain.

The reparametrization mode can also lead to correction for self-energy in the $1/M$ order as we have mentioned in the previous section. 
To see that, we notice the coupling to chaotic Majorana also mediate quadratic fluctuation of infinitesimal reparametrizations, i.e.,
\bea
	\delta G^p_\chi(\t_1,\t_2) &=& G^p_\chi(\tau_{12})  [ B(\tau_1, \tau_2) + C(\tau_1, \tau_2)], \\
	C(\tau_1, \tau_2) &=&  \frac{2p}q \left( \big( \frac{\epsilon(\tau_1)-\epsilon(\tau_2)}{2\sin \frac{\tau_{12}}2} \big)^2 - \frac12 \big( \epsilon'(\tau_1) + \epsilon'(\tau_1)^2\big)\right) \\
	&&  + \frac12 \big( \frac{2p}q \big)^2 \left(\epsilon'(\tau_1) + \epsilon'(\tau_1)- \frac{\epsilon(\tau_1)-\epsilon(\tau_2)}{\tan \frac{\tau_{12}}2} \right)^2.
\eea
The quadratic fluctuation leads to the self-energy correction
\bea
	\Sigma^{(1)}(\tau_1, \tau_2) = \frac{\J}{M \alpha_S} \Big(\frac1{2\sin \frac{\tau_{12}}2} \Big)^{\frac{2p}q} \Big( \frac{2p}q  \frac{\t_{12}^2 - 2\pi \t_{12}+2 - 2 \cos \t_{12} + 2(\pi - \t_{12}) \sin \t_{12}}{(2 \sin \frac{\t_{12}}2)^2} \nn
	 \\ + \frac12 \big( \frac{2p}q \big)^2 \big( -2 + \frac{\t_{12}}{\tan \frac{\t_{12}}2} \big) \big( -2 + \frac{\t_{12}-2\pi}{\tan \frac{\t_{12}}2} \big) \Big).
\eea
It is easy see this is not only suppressed by $1/M$, but also subleading at the conformal limit $\beta \rightarrow \infty$ because it is proportional to $\beta^{-2p/q}$ once the temperature is restored in the expression.

\subsection{Backreaction: a large $q$ study} \label{subsec:backreaction}

In this section, we develop a large $q$ study to investigate the backreaction of the local Majorana on the chaotic Majorana. 
We are going to use parameters suitable for the large $q$ expansion. 
The interaction among the chaotic Majorana, and the coupling between the chaotic Majorana and the local Majorana are defined by $\J$ and $\V$ through the second equality in~(\ref{eq:J2}) and~(\ref{eq:V2}). 
Although we still need to take the large $N$ limit before the large $q$ limit, we can consider $M=N$ and the backreaction is controlled by the large $q$ expansion.

For the chaotic Majorana, the large-$N$ action $I_C$ is the same as the original SYK model. 
By making the ansatz, $G(\tau_1, \tau_2) = \frac12 \sgn(\t_1 - \t_2) [ 1 + \frac1q g(\tau_1, \tau_2) ]$ (see Appendix~\ref{append:large-q}), we get the following large-$q$ effective action,
\bea
	- \frac{I_C}{N} &=& \frac1{q^2} \int d\t_1 d\t_2 \left( - \frac1{16}  \partial_{\t_1} g \partial_{\t_2} g + \frac{\J^2}{4}  e^{g(\tau_1,\tau_2)} \right).
\eea

The coupling between the local and the chaotic Majorana is
\bea
	- I_{LC}= \frac{\V^2}{4p^2} \int d\t_1 d\t_2  (2G_\chi(\tau_1, \tau_2))^p \sum_r \psi_r(\tau_1) \psi_r(\tau_2).
\eea
We will set $p = \kappa q$, and take the large $q$ limit while keep $\kappa$ fixed. Because the local Majorana is a quadratic theory, we can first integrate it out. The coupling term becomes
\bea
	\langle e^{-I_{LC}} \rangle &=& \left\langle e^{ \frac{\V^2}{4p^2} \int d\t_1 d\t_2  (2G_\chi(\tau_1, \tau_2))^p \sum_r \psi_r(\tau_1) \psi_r(\tau_2) } \right\rangle \\
	&=&  e^{ \frac{\V^2}{4p^2} \int d\t_1 d\t_2  (2G_\chi(\tau_1, \tau_2))^p \sum_r \langle \psi_r(\tau_1) \psi_r(\tau_2) \rangle },
\eea
where the bracket denotes the path integral over local Majorana fermion with action $I_L$, and in the second line, we have used the following property, 
\bea
	\Big \langle \big[\sum_r \psi_r(\tau_1) \psi_r(\tau_2) \big]^n \Big \rangle =\big[\sum_r \langle \psi_r(\tau_1) \psi_r(\tau_2) \rangle\big]^n + \mathcal O(N^{n-1}).
\eea
This property arises because the contribution from the correlation for $r \ne r'$ is exponentially suppressed by $ e^{- \frac{\pi |r-r'|}{w \beta}}$, and is thus suppressed by $N^{-1}$.
For instance, at $n=2$, besides the term proportional to $N^2$, there is another term as below that is suppressed by $1/N$,
\bea
	\sum_{r,r'} G_{\psi,0}(\tau_{12}, r-r') G_{\psi,0}(-\tau_{12}, r'-r) \approx N \int dr e^{- \frac{2\pi |r|}{w \beta}} = N \frac{w \beta}\pi < N^2.
\eea
The inequality holds true because $N$ is the largest parameter in the problem.

Thus, the effect of the local Majorana is the following contribution to the effective action, 
\bea
	\langle e^{- I_{LC}} \rangle &\approx& e^{\frac{\V^2}{4p^2} \int d\t_1 d\t_2  (2G_\chi(\tau_1, \tau_2))^p \sum_r \langle \psi_r(\tau_1) \psi_r(\tau_2) \rangle} \\
	&=& e^{ N \frac{\V^2}{4p^2} \frac1{\pi v} \int d\t_1 d\t_2  \frac{\pi}{\beta \sin \frac{\pi \tau_{12}}\beta} (2G_\chi(\tau_1, \tau_2))^p } \\
	&\approx & \exp \left( N \frac{\V^2}{4\kappa^2 q^2} \frac1{\pi v} \int d\t_1 d\t_2  \frac{\pi}{\beta \sin \frac{\pi |\tau_{12}|}\beta} e^{\kappa g(\tau_1, \tau_2) } \right).
\eea
In the last step we take the large $q$ limit while keep $p/q= \kappa $ fixed. Combined with this term, the large-$q$ action for the chaotic Majorana reads,
\bea
	- \frac{\tilde I_C}{N} &=& \frac1{8q^2} \int d t_1 d t_2 \left( - \frac1{2}  \partial_{t_1} g \partial_{t_2} g - 2\J^2  e^{g(t_1,t_2)}  - \frac{2\V^2}{\kappa^2 v} \frac{1}{\beta \cosh \frac{\pi t_{12}}\beta} e^{\kappa g(t_1, t_2) }\right),
\eea
where we have analytically continued to real times. We review the large $q$ action and the way to extract the OTOC in Appendix~\ref{append:large-q}. Assuming $g$ depends on the time difference only, the equation of motion reads
\bea
	\partial^2_t g(t) = - 2\J^2 e^{g(t)} - \frac{2\V^2}{\kappa^2 v} \frac{1}{\beta \cosh \frac{\pi t}\beta} e^{\kappa g(t) } .
\eea
Without the coupling, namely $\V = 0$, the solution reads
\bea
 g(t)= \log \frac{\alpha^2}{\J^2 \cosh^2\alpha t }, \quad \alpha = \J \sin \gamma, \quad \gamma = \frac\pi2 - \frac{\alpha\beta}2.
\eea
With the coupling, we are not able to solve the equation. 
In order to make an estimate, we can approximate the correlation function from local Majorana as a delta function, $\frac1{\cosh \frac{\pi t}\beta} \approx \beta \delta(t) $, that would be a better approximation at higher temperature.
Equivalently, it is exactly the case when one considers the Brownian-type couplings~\cite{Saad:2018semiclassical, Sunderhauf:2019quantum, Jian:2020note, jian2021phase}.
Nevertheless, we expect that the essential physics does not depend on it.
Then the equation of motion and the solution are
\bea
	\partial^2_t g(t) = - 2\J^2 e^{g(t)} - \frac{2\V^2}{\kappa^2 v} \delta(t) e^{\kappa g(t)}, \quad \tilde g(t) =2 \log \frac{\alpha}{\J \cosh\alpha t } - \left( \frac{\alpha}{\J} \right)^{2\kappa} \frac{\V^2 |t|}{\kappa^2 v} .
\eea

The coupling to the local Majorana at high temperature leads to a correction to the saddle-point solution. 
We are now interested in how it affects the chaotic behavior. 
The channel through the chaotic Majorana given in~(\ref{eq:v4}) is the propagator of the bilocal field $G_\chi(t,t')$. 
In the large $q$ expansion, it reduces to the correlation function of $g(t, t')$. 
We can expand the effective action to the quadratic order $g(t_1,t_2) = \tilde g(t_{12}) + \delta g(t_1, t_2)$ to get its correlation function,
\bea
	- \frac{\tilde I_C}{N} &=& \frac1{8q^2} \int \delta g \left( \frac14 \partial_{\bar t_{12} }^2 - \partial_{t_{12}}^2 - 2\J^2  e^{\tilde g(t_{12})}  - \frac{2\V^2}{\kappa^2 v} \frac{1}{\beta \cosh \frac{\pi t_{12}}\beta} e^{\kappa \tilde g(t_{12}) }\right)\delta g.
\eea
where $\bar t_{12} = \frac{t_1 + t_2}2$ and $t_{12}= t_1 - t_2$. 
We then assume $\delta g(t_1, t_2) = e^{\lambda \bar t_{12}}\psi_\lambda(t_{12})$ to get the long time behaviors, where $\lambda$ is the Lyapunov exponent. Plugging the ansatz into the effective action, the Lyapunov exponent is determined by the one dimensional Hamiltonian $( \frac{\lambda^2}4 + H(t) ) \psi_\lambda(t) = 0$, where
\bea
	H(t) = - \partial_t^2 - \frac{2\alpha^2}{\cosh^2 \alpha t} e^{- \left( \frac{\alpha}{\J} \right)^{2\kappa} \frac{\V^2|t|}{\kappa^2 v} } - \frac{2\V^2}{\kappa^2 v} \frac{1}{\beta \cosh \frac{\pi t}\beta} \left( \frac{\alpha}{\J\cosh \alpha t} \right)^{2\kappa} e^{- \left( \frac{\alpha}{\J} \right)^{2\kappa} \frac{\V^2|t|}{\kappa v} }.
\eea
Although we are not able to solve the Hamiltonian, we treat $\V$ as a perturbation. 
The unperturbed Hamiltonian for $\V=0$ has a bound state $\psi_{2\alpha}(t) = \sqrt{\frac{\alpha}{2}} \frac1{\cosh \alpha t}$ leading to the well known Lyapunov exponent $\lambda = 2\alpha$~\cite{Maldacena:2016remarks} in the large-$q$ limit. 
We can get the first order perturbation by using the zeroth order wave function, namely 
\bea
	\int_{-\infty}^\infty dt \psi_{2\alpha}^*(t) H(t) \psi_{2\alpha}(t) &\approx& - \left(\alpha^2 - \alpha \frac{\V^2 }{\kappa^2 v} \left( \frac{\alpha}{\J} \right)^{2\kappa} \left(  \frac13 \log(\frac{16}e)   + C \right) \right) + \mathcal O(\V^3), \\
	C &=& -   \int_{-\infty}^\infty  \frac{dt}\beta \frac1{\cosh \frac{\pi t} \beta} \left(\frac1{\cosh \alpha t} \right)^{2\kappa + 2}.
\eea
The effect of the local Majorana fermion is to give a correction to the Lypunov exponent,
\bea
	\lambda = 2\alpha - \frac{\V^2 }{\kappa^2 v} \left( \frac{\alpha}{\J} \right)^{2\kappa} \left(  \frac13 \log\left(\frac{16}e\right)   + C \right) + \mathcal O(\V^3).
\eea
This shows that for $M=N$, the chaotic Majorana receives a correction from the coupling to the local Majorana but remains chaotic as long as the coupling is not too large.

At the conformal limit, we can get an analytic result $C = - \frac{\Gamma(\frac32 + \kappa)}{\sqrt \pi \Gamma(2 + \kappa)} $, and 
\bea
\lambda = \frac{2\pi}\beta \left( 1  - \frac{\V^2 }{\kappa^2 v} \left( \frac{\pi}{\beta\J} \right)^{2\kappa} \left(  \frac13 \log(\frac{16}e)   - \frac{\Gamma(\frac32 + \kappa)}{\sqrt \pi \Gamma(2 + \kappa)}\right) \right).
\eea
The correction is negative even for $\kappa \rightarrow 1$, but in fact, this correction is subleading in temperatures because the first order correction to $2\pi/\beta$ is from $\alpha \approx \frac{\pi}\beta ( 1 - \frac{2}{\beta \J} )$~\cite{Maldacena:2016remarks}.
Though we have made assumptions that rely on high temperatures, the conformal approximation could serve as a consistency check that the approximation does not violate the chaos bound~\cite{Maldacena:2015a}. 
We have also considered an analysis in the conformal limit in Appendix~\ref{append:conformal}, where the Brownian-type approximation is implemented. 
The results show that the backreaction from the local Majorana is irrelevant.

\section{Entanglement dynamics after a global quench} \label{sec:quench}

As we have seen from the OTOC, owing to the coupling to the chaotic Majorana system, operators spread exponentially fast in the local Majorana chain. 
This seemingly non-local structure is nonetheless invisible in local probes. In this section, we will show that the full coupled $LC$ system is indeed a fast scrambler~\cite{Lashkari:2011towards} that takes a much shorter time to scramble than the local Majorana alone. 
More explicitly, starting from a product state, the R\'enyi entropy between two halves grows as the state evolves, and we show that the rate of increase of the R\'eny entropy is proportional to the system size. 
This is a signature of a fast scrambler as all degrees of freedom participate in scrambling the information. By contrast, without coupling to the chaotic Majorana the rate of increase of entanglement between two halves of the chain will be a constant independent of the system size because the entanglement is carried by quasiparticle pairs that propagate at a constant speed.

\subsection{Quench protocol and setup} \label{subsec:setup}

We introduce the physical setup including the quench protocol, and formulate the quantity that we will calculate. The time evolution operator is generated from the Hamiltonian
\bea
	U(T) = \T e^{-\ii \int_0^T ds H(s) ds }, 
\eea
where $\T$ denotes the time ordering, and $T$ is the evolution time. 
Consider an initial pure state (unnormalized) $|\Psi \rangle$, that is not an eigenstate of the Hamiltonian, after $T$ evolution, the density operator becomes
\bea
	\rho(T)= \frac{|\Psi(T) \rangle \langle \Psi(T) |}{Z(T)}, \quad | \Psi(T) \rangle = U(T) |\Psi \rangle, \quad Z(T) = \langle \Psi | U^\dag(T) U(T) | \Psi \rangle.
\eea
The function $Z(T)$ is to make sure that the density operator is properly normalized.
In the case of unitary evolution, it is time independent, $Z(T) = \langle \Psi | \Psi \rangle$.

In the following, we will focus on the average purity (and the second R\'enyi entropy), 
\bea
	P(T)= \overline{\Tr[ \rho(T)^{\otimes 2} \S]}, \quad \S = \text{SWAP}.
\eea
Note that the swap operator acts on the subregion whose purity is to be calculated. 
In principle, we need to deal with the disorder average in the denominator, which involves a replica trick. 
Here, we assume the quantity is self-averaged, and we calculate
\bea
	P(T) = \frac{\overline{Z_2(T) }}{\overline{Z(T)^2}}, \quad Z_2(T) = \Tr[(| \Psi (T)\rangle \langle \Psi(T) |)^{\otimes 2} \S], \quad Z(T)^2 = \Tr[(| \Psi(T) \rangle \langle \Psi(T) |)^{\otimes 2}].
\eea
The difference between the numerator and denominator is the boundary condition induced by the swap operator.

To construct the initial state, we double the system to the left and the right side, and consider the tensor product of EPR states between the left and the right system~\cite{Gu:2017spread, jian2021phase}. 
For each pair of Majorana operators, including both the pair of the local Majorana $\psi_{r}^L$ and $\psi_{r}^R$, and the pair of the chaotic Majorana $\chi_{j}^L$ and $\chi_{i}^R$, the EPR state is defined by
\bea
	&& (\psi_r^L + \ii \psi_r^R) | \psi_r, \text{EPR} \rangle = 0, \quad \forall r=1,...,N. \\
	&& (\chi_j^L + \ii \chi_j^R) | \chi_j, \text{EPR} \rangle =0, \quad \forall i=1,...,M.
\eea
The initial state is the tensor product of these EPR states, i.e.,
\bea
	|\Psi \rangle = \left(\otimes_{r=1}^{N} | \psi_r, \text{EPR} \rangle \right) \otimes \left(\otimes_{j=1}^{ M} | \chi_j, \text{EPR} \rangle \right).
\eea

\begin{figure}
    \centering
    \includegraphics[width=0.45\textwidth]{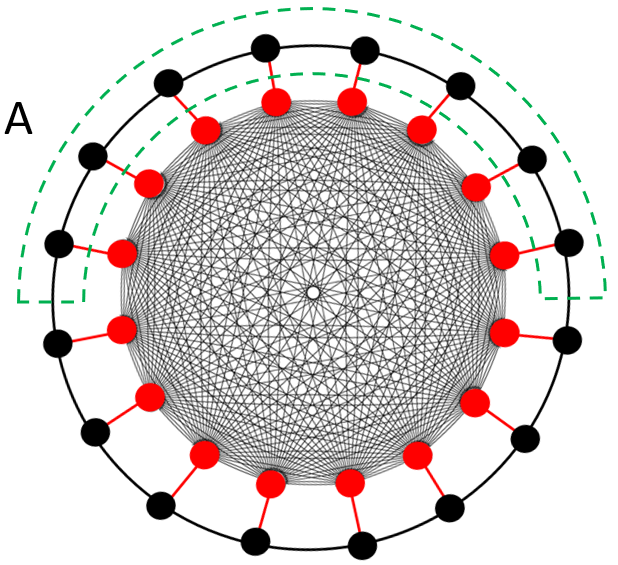}
    \caption{A schematic plot of regions between which we choose to calculate the second R\'enyi entropy in the main text. The region inside the green dashed line is region $A$, and the compliment is region $B$. If the local Majorana do not couple to the chaotic Majorana, any information transmission between two regions must travel through the boundary and the speed of it is given by the quasiparticle velocity. Whereas, if the coupling is nonzero, non-local information can transmit much faster with the help of the chaotic Majorana.}
    \label{fig:regions}
\end{figure}

Although the entanglement between the left and right side is maximal for this initial state, there is {\it no} entanglement between different Majorana fermions in the same side. 
Moreover, this state is not an eigenstate of the Hamiltonian, and will evolve nontrivially. 
We divide the system into two subregions, $A$ and $B$ as shown in Fig.~\ref{fig:regions}, and study the time evolution of the second R\'enyi entropy between them. 
Region $A$ contains half of the local Majorana chain, $\psi_r$, $i=r,...,N/2$, and region $B$ is the compliment of region $A$ that contains $\psi_r$, $r=N/2,...,N$ and $\chi_j$, $j=1,...,M$\footnote{We assume $N$ is an even integer}.
Note that the chaotic Majorana are all located in region $B$. 
If there is no coupling between local Majorana and chaotic Majorana, $H_{LC}=0$, what we calculate is the R\'enyi entropy between two halves of the local Majorana chain since the time-evolved state maintains a tensor produce between two types of Majorana $\psi$ and $\chi$, i.e.,
\bea
	|\Psi (T) \rangle =\left( \T e^{-\ii \int_0^T ds H_L(s) ds }  \otimes_{i=r}^{ N}  | \psi_r, \text{EPR} \rangle\right) \otimes \left( \T e^{-\ii \int_0^T ds H_C(s) ds } \otimes_{j=1}^{ M} | \chi_j, \text{EPR} \rangle  \right).
\eea
On the other hand, we can turn on the coupling between the local Majorana and the chaotic Majorana to investigate how the chaotic Majorana changes the entanglement dynamics of the local Majorana chain.

\begin{figure}
	\centering
	\includegraphics[width=0.5\textwidth]{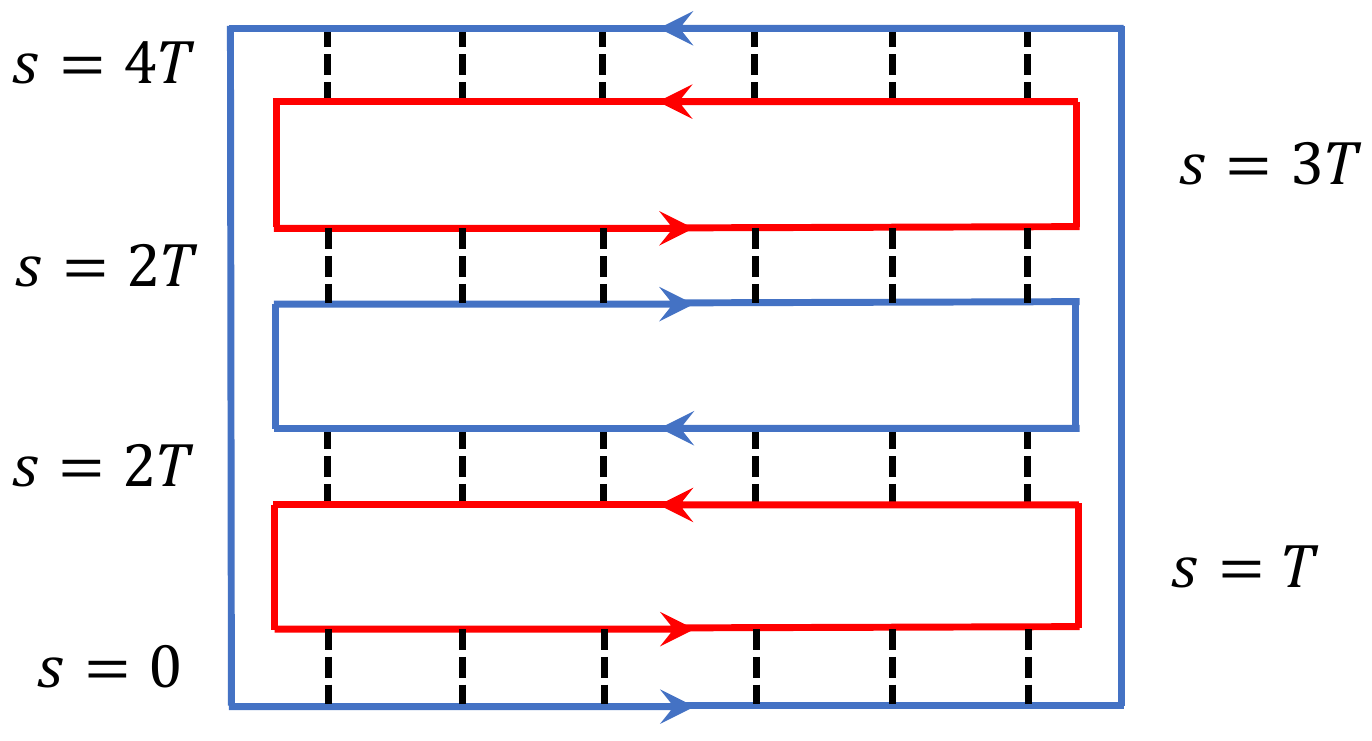}
	\caption{\label{fig:contour} A schematic labeling of the contour. The blue (red) line indicates the subregion A (B). The dashed line indicates couplings between the two regions. The arrow denotes the real time (unitary) evolution.}
\end{figure}

We use path integral to calculate the $\overline{Z_2(T)} $ and $\overline{Z(T)^2}$. 
We use parameter $0<s<2T$ ($2T<s<4T$) to denote the first (second) replica~\cite{Chen:2020replica,zhang2020entanglement,Jian:2020note}. 
In each replica, there are two Keldysh contours distinguished by $2n T <s< 2n T +T$ and $2n T +T <s< 2n T +2T$, where $n=0,1$ correspond to the two replicas, respectively. 
A schematic labeling of the contour is shown in Fig.~\ref{fig:contour}, where the blue (red) line indicates the subregion $A$ ($B$). 
Since the two regions are distinct, we introduce two bilocal fields for the local Majorana $\psi_r$,
\bea 
	G_{\psi,A} (s_1, s_2)= \frac2{N} \sum_{r=1}^{N/2} \psi_r(s_1) \psi_r(s_2), \quad G_{\psi,B} (s_1, s_2)= \frac2{N} \sum_{r=N/2+1}^{N} \psi_r(s_1) \psi_r(s_2).
\eea
While for the chaotic Majorana $\chi_j$, one bilocal field $G_{\chi, B}= \frac1M \sum_{j=1}^M \chi_j(s_1) \chi_j(s_2)$ is enough because they all are located in the same region $B$. 

As usual, it is convenient to introduce another bilocal field, the self-energy, to enforce the $G$ fields as the propagators of the Majoranas.
In terms of the bilocal fields and the corresponding self-energies, the path integral representation of the purity~\cite{Penington:2019replica,Chen:2020replica, zhang2020entanglement, Jian:2020note} is $P(T) = \frac{\overline{Z_2(T)}}{\overline{Z(T)^2}}$, and
\bea 
    \overline{Z_2(T)} = \int D G_{\psi, A} D\Sigma_{\psi, A} D G_{\psi, B} D\Sigma_{\psi, B}D G_{\chi, B} D\Sigma_{\chi, B} e^{- I_2}, \\
    \overline{Z(T)^2} = \int D G_{\psi, A} D\Sigma_{\psi, A} D G_{\psi, B} D\Sigma_{\psi, B}D G_{\chi, B} D\Sigma_{\chi, B} e^{- I_0}.
\eea
We define the following action 
\bea
	&& -I(F_{\psi,A}, F_{\psi,B}, F_{\chi,B})   = \frac12 \Tr \log \left( \left( \ba{cccc} F_{\psi,A} - \Sigma_{\psi, A} & 0 \\ 0 & F_{\psi,B} - \Sigma_{\psi,B} \ea \right) - \ii w f(s_1) \delta(s_1-s_1) h_{r_1,r_2} \right) \nn\\
	&& + \frac{M}2 \Tr \log(F_{\chi,B}^{-1} - \Sigma_{\chi,B}) - \frac{N}4 \int (\Sigma_{\psi,A} G_{\psi,A} + \Sigma_{\psi, B} G_{\psi,B})  - \frac{M}2 \int \Sigma_{\chi, B} G_{\chi,B} \nn\\
	&& + \frac{M J^2}{2q} \int f(s_1) f(s_2) (G_{\chi, B})^q + \frac{NV^2}{2p} \int f(s_1)f(s_2) (G_{\chi, B})^p \frac{G_{\psi,A} + G_{\psi,B}}2,
\eea
where the function $f(s) = -\ii$ for $s \in (0, T) \cup (2T, 3T)$ and $f(s)= \ii$ for $s \in (T,2T) \cup (3T, 4T)$ is to capture the time evolution in the Keldysh contours. 
The first (second) $\Tr \log$ term is resulted from integrating over the local (chaotic) Majorana $\psi_r$ ($\chi_j$).
The trace that involves the local Majorana $\psi$ contains trace in both the $s$ and the $r$ space, while the trace that involves the chaotic Majorana $\chi$ contains only $s$ space. 
The two-by-two matrix in the $\Tr \log$ term for the local Majorana corresponds to the two half chains of the local Majorana separating two regions.
To incorporate the different boundary conditions, the $F$ function is chosen differently. We define
\bea
    (F^{(0)})^{-1}(s,s') = \frac12 \sgn(s-s'), \quad s,s' \in (0,2T) \text{ or } s,s' \in (2T,4T), \\
    (F^{(1)})^{-1}(s,s') = \frac12 \sgn(s-s'), \quad s,s' \in (T,3T) \text{ or } s,s' \in (0,T)\cup(3T,4T).
\eea
In $\overline{Z(T)^2}$, there is no twist boundary condition, while in $\overline{Z_2(T)}$, the twist boundary condition is implemented in subregion $A$. To account for these boundary conditions, the actions are given respectively,
\bea
    I_0 = I(F^{(0)}, F^{(0)}, F^{(0)}), \quad I_2 = I(F^{(1)}, F^{(0)}, F^{(0)}).
\eea

In the following, we will take $N=M \gg 1$ and use the saddle-point approximation to evaluate the R\'enyi entropy.

\subsection{Time evolution of R\'enyi entropy} \label{subsec:evolution}

From $I_2$, by varying the bilocal fields, the equation of motion for the local Majorana reads
\bea
\label{eq:local_eomA}	G_{\psi,A} &=& \frac2N \Tr\left[ \left( \left( \ba{cccc} F^{(1)} - \Sigma_{\psi, A} & 0 \\ 0 & F^{(0)} - \Sigma_{\psi,B} \ea \right)^{-1} - \ii w f h \right) \cdot \left( \ba{cccc} 1 & 0 \\ 0 & 0 \ea \right) \right], \\
\label{eq:local_eomB}	G_{\psi,B} &=& \frac2N \Tr \left[ \left( \left( \ba{cccc} F^{(1)} - \Sigma_{\psi, A} & 0 \\ 0 & F^{(0)} - \Sigma_{\psi,B} \ea \right)^{-1} - \ii w f h \right). \left( \ba{cccc} 0 & 0 \\ 0 & 1 \ea \right) \right], \\
	\Sigma_{\psi,A} &=& \Sigma_{\psi,B} = \frac{V^2}{p} f_1(s_1) f_2(s_2) G_{\chi,B}^p, 
\eea
where the trace in the first two equations acts on the $r$ space, and for the chaotic Majorana reads
\bea
	G_{\chi, B} &=& (F^{(0)} - \Sigma_{\chi, B})^{-1}, \\
	\Sigma_{\chi, B} &=& J^2 f(s_1) f(s_2) G_{\chi,B}^{q-1} + \frac{N}M V^2 f_1(s_1) f_2(s_2) G_{\chi,B}^{p-1} \frac{G_{\psi,A} + G_{\psi,B}}2.
\eea
We do not present the equation of motion from $I_0$ which reduces to the single contour case.

We numerically solve the equation of motion by iterations~\cite{Maldacena:2016remarks, Chen:2020replica, Jian:2020note}. 
For simplicity, we set $N=M$. 
Because of the presence of the hopping matrix $h$, the factor $N$ does not drop out from the equation of motion in~(\ref{eq:local_eomA}, \ref{eq:local_eomB}) unlike the regular SYK model, instead, the $h$ matrix is $N \times N$. 
So in the calculation, we set $N$ to be a finite number to iteratively search for the saddle-point solution. 
Once we have gotten the saddle-point solution, we plug it back to the action to get the purity for a finite $N$. 
We will calculate the purity for different $N$ and extrapolate the result for $N \rightarrow \infty $.

If there is no coupling to the chaotic Majorana, the local Majorana chain is a free model whose entanglement is carried by entangled pairs of quasiparticles that are excited from the global quench~\cite{Calabrese:2005evolution}. 
Namely, the initial state has a finite energy density with respect to the post-quench Hamiltonian $H$ and the global quench excites entangled pairs of quasiparticles. 
When one of the entangled quasiparticles moves across the boundary between region $A$ and $B$, the R\'enyi entropy grows by a constant amount. 
So one expects the R\'enyi entropy grows linearly at a speed of twice the quasiparticle velocity. 
The factor two is due to the two boundaries between region $A$ and $B$ of the local Majorana. 

\begin{figure}
	\centering
	\subfigure[]{\label{fig:s2_linear}
		\includegraphics[width=0.4 \textwidth]{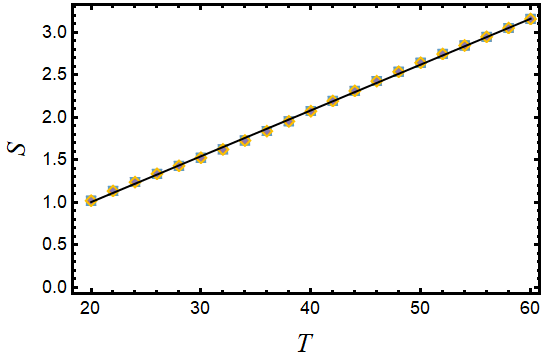}} \qquad \qquad
	\subfigure[]{\label{fig:s2_slope}
		\includegraphics[width=0.4 \textwidth]{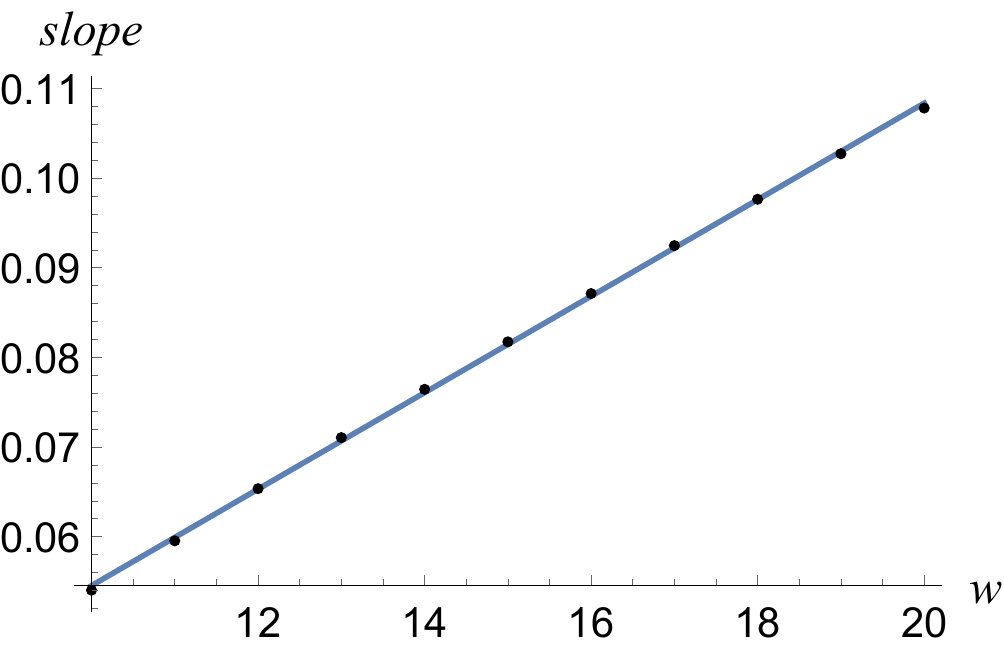}}
	\caption{(a) The time evolution of R\'enyi entropy of region $A$ without coupling to the chaotic Majorana for $N=18,20,...,32$ for $w=10$. It shows a single line because all the lines collapse into each other. (b) The fitted slope of time evolution of the R\'enyi entropy as a function of the hopping amplitude. We take $w=10,...,20$ and $T \in (20,60)$. The parameters are $\beta = 0$, $J=1$ and $V=0$. We discretize $T$ by $40$ points.}
\end{figure}

Figure~\ref{fig:s2_linear} shows the time evolution of the R\'enyi entropy of half of the local Majorana chain without coupling to the chaotic Majorana. 
The linearity of growth of the entropy is clear. 
In the figure, we plot the R\'eny entropy for $N=18, 20,..., 32$ where the region $A$ has $N/2=9, 10,...,18$ sites, but they all collapse into a single curve because the increasing rate does not depend on the size of region $A$. 
The slope is determined by the quasiparticle velocity and, in our case, it is proportional to the hopping amplitude $w$. 
We also check it in Fig.~\ref{fig:s2_slope}, where the slope is proportional to $w$ as shown.

\begin{figure}
	\centering
	\subfigure[]{\label{fig:s2_single}
		\includegraphics[width=0.4 \textwidth]{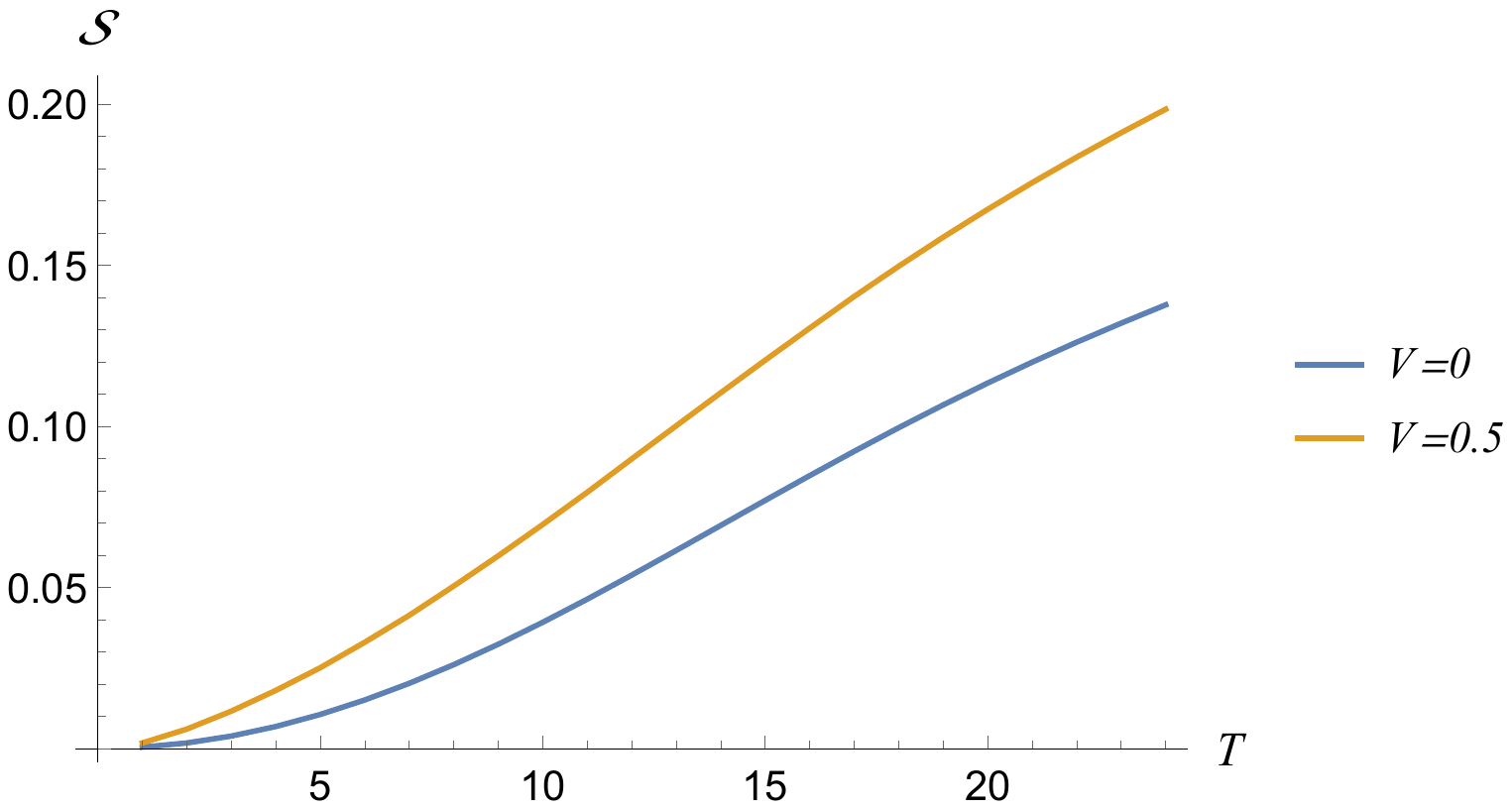}} \\
	\subfigure[]{\label{fig:s2_extrapolation}
		\includegraphics[width=0.475 \textwidth]{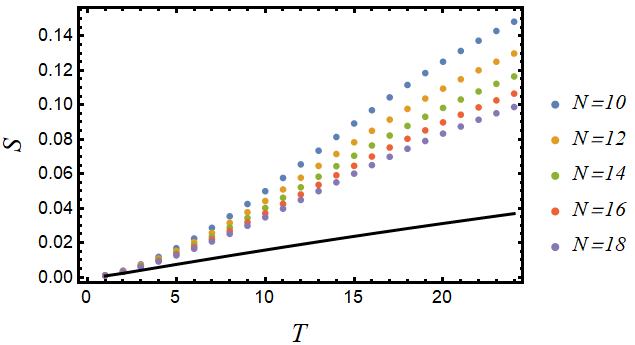}} \quad
	\subfigure[]{\label{fig:extrapolation} 
		\includegraphics[width=0.4 \textwidth]{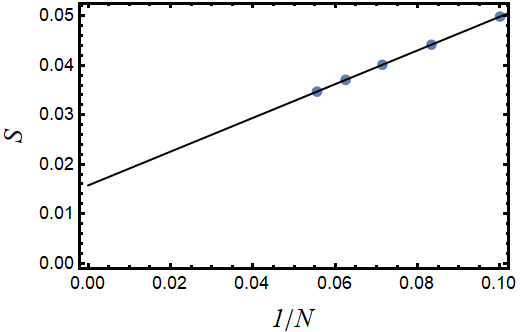}}
	\caption{(a) The time evolution of the R\'enyi entropy per Majorana fermion with and without coupling to the chaotic Majorana for $N=18$. (b) The time evolution of R\'enyi entropy per Majorana fermion for $N=10,...,18$ and $V=0.5$ are shown by dots and its extrapolation to $N \rightarrow \infty$ is the black line. The parameters are $\beta=0$, $J=1$ and $w=10$. We discretize $T$ by $40$ points. (c) An example of the extrapolation procedure. The plot is shown for $T=10$.}
\end{figure}

Since the rate increase of the R\'enyi entropy is independent of the size of the region $A$, the time to reach the equilibrium entropy is proportional to $N$. 
If we look at the R\'enyi entropy per Majorana, the rate scales down with the size of the region $A$ since scrambling time is proportional to $N$. 
In other words, the rate of entropy increase per Majorana is $v/N$, where $v$ is the quasiparticle velocity, and it scales to zero in the $N \rightarrow \infty$ limit. 

Now we turn on the coupling $V>0$ between the local Majorana and the chaotic Majorana. 
Figure~\ref{fig:s2_single} shows that for a fixed $N=18$, the rate of increase is enhanced by the coupling $V$. 
The effect is more dramatic if we scale $N \rightarrow \infty$ and look at the R\'enyi entropy per Majorana. 
Without coupling to the chaotic Majorana, the rate extrapolates to zero at large $N$. However, when the coupling is nonzero, the extrapolation to $N \rightarrow \infty $ of the rate of increase remains finite as shown by the black line in Fig.~\ref{fig:s2_extrapolation}.  
This is in some sense the net result from the coupling to the chaotic Majorana, since taking the limit $N \rightarrow \infty$ eliminates the contribution from the entangled quasiparticles. 
The finite slope for the entropy per Majorana in the $N \rightarrow \infty $ limit indicates that the scrambling time is no longer proportional to $N$. 
Actually, since the slope is finite, this naively indicates that the scrambling time is a constant independent of $N$. 
But this cannot be true as we expect the slope to decrease at larger time and finally to lead to a $\log N$ scrambling time~\cite{Lashkari:2011towards}. An example of the extrapolation procedure is shown in Fig.~\ref{fig:extrapolation} for $T=10$, and the black line in Fig.~\ref{fig:s2_extrapolation} is obtained by extrapolating data at each $T$.

To conclude, we have constructed a solvable model and have shown that although the simple information in the local Majorana remains local, the coupled system is a fast scrambler where non-local information can spread much faster than the simple signal.

\section{A black hole puzzle} \label{sec:BH}

In this section, we apply the results of previous sections to a puzzle about black hole dynamics. We first recall the setup discussed by Shor~\cite{Shor:2018scrambling}. 
Then we formulate the puzzle for AdS black holes of roughly AdS radius size, and argue that there is a sharp problem. Finally, we sketch a resolution of the problem based on the idea that the near horizon dynamics of the black hole is inherently non-local.

\subsection{Shor's cell model}

Consider a Schwarzschild black hole in 3+1 spacetime dimensions. The metric in Schwarzschild coordinates is
\beq
ds^2 = - \left( 1 - \frac{2 G M}{r}\right) dt^2 + \left( 1 - \frac{2 G M}{r}\right)^{-1} dr^2 + r^2 d\theta^2 + r^2 \sin^2 \theta d\phi^2.
\eeq
Henceforth we set Newton's constant $G=1$ and generally use natural units. The radius of the black hole is thus $r_s =2 M$, the entropy is $S \sim M^2$, the energy is $M$, and the temperature\footnote{In this section, we use $T=1/\beta$ to denote the temperature.} is $T \sim 1/M$. We have in mind that $M$ is very large.

To make a simple computational model of the near horizon region, Shor considers partioning the spacetime into cells with the defining property that the Schwarzschild time required for light to cross a cell is of order $M$. From the point of view of the black hole as a chaotic quantum system at temperature $T \sim 1/M$, this time is simply the thermal time $\beta =1/T \sim M$. In essence, we are viewing the black hole a network of computational cells such that any cell can communicate with its nearest neighbor in a Schwarzschild time of order $\beta$.

The rough size of the cells is determined as follows. Let $h= r-2M$ denote the height of the cell above the horizon. Everything will be discussed to leading non-trivial order in $h$ and ignoring order one constants. Let $\Delta x$ and $\Delta r$ denote the width of the cell perpendicular to and along the horizon, respectively. The parallel width, which we may think of as $r \Delta \theta$ with $\Delta \theta$ an angular size, satisfies
\beq
\Delta x \sim  \sqrt{\frac{h}{M}} M = \sqrt{M h}.
\eeq
The perpendicular width satisfies
\beq
\sqrt{\frac{M}{h}} \Delta r \sim  \sqrt{\frac{h}{M}} M
\eeq
or
\beq
\Delta r \sim h.
\eeq
Note that the proper radial width is $\sqrt{M/h} \Delta r \sim \sqrt{M h} \sim \Delta x$.

At some cutoff height $h = \alpha M$ with $\alpha$ order one (Shor takes $\alpha = 1$ corresponding to the photon sphere), the number of cells is order one. 
The number of cells at height $h$ is approximately $\left( \frac{M}{\sqrt{Mh}}\right)^2 \sim \frac{M}{h}$. 
To count the total number of cell layers, note that the cell height increases exponentially away from the horizon. 
Starting from a cell of height $h_0$, the radial width is also $\Delta r = h = h_0$, so the next cell has height $h_1 = 2h_0$. 
Its radial width with $\Delta r = h = 2h_0$, so the next cell has height $h_2 = 4 h_0$. 
Starting from near the horizon at $h=h_0$, $i$-th cell away from the horizon has height $2^i h_0$. Hence the number of cells needed to reach from height $h_0$ to height $M$ is of order $\log \frac{M}{h_0}$. 
It is natural to take $h_0 \sim 1/M$ corresponding to cells of Planck-scale proper size, $\Delta x \sim 1$. The cells closest to the horizon form the stretch horizon; there are order $S \sim M^2$ of them arranged in a 2D array on the spherical horizon.

We comment that the cell model is essentially a discrete version of an older construction known as the ``optical metric'' which consists of taking the full metric and dividing by $g_{tt}$ to produce an effective spatial metric in which light propagates with unit speed. 
For the near horizon region of a black hole, we can use Rindler coordinates to write $ds^2 = - \rho^2 d\eta^2 + d\rho^2 + d \vec{x}^2$ where $\rho \sim \sqrt{M h}$ is the proper distance to the horizon and $\eta \sim t/M$ is a rescaled time variable. 
The optical metric is then
\beq
ds^2_{\text{opt}} = -d\eta^2 + \frac{d\rho^2 + d\vec{x}^2}{\rho^2},
\eeq
which corresponds to light moving in hyperbolic space. This gives another point of view on the exponentially increasing number of cells towards the horizon ($\rho=0$).

To complete the model, we take each computational cell to contain an order one number of qubits. This is justified by estimating the entropy of quantum fields around the black hole. Viewing the near horizon region as Rindler-like, quantum fields in the vicinity of the black hole can be regarded as being in a thermal state with a position dependent temperature $T_{\text{loc}}(r) = \frac{T}{\sqrt{g_{tt}}}\sim \frac{1}{\sqrt{hM}}$. When this temperature is higher than a relevant mass-scale, we can treat the field as scale-invariant and estimate the thermal entropy density as $T_{\text{loc}}^3$. The entropy in one cell is thus
\begin{equation}
    S_{\text{cell}} \sim T_{\text{loc}}^3 (\Delta x)^2 \sqrt{\frac{M}{h}} \Delta r \sim 1.
\end{equation}
The dynamics of course depends on the character of the fields and their interactions, but we don't expect the entropy density to vary dramatically from weak to strong coupling. Hence, we estimate that each cell contains order one bit of entropy per (effectively) massless field. These are assumed to correspond to active quantum degrees of freedom that we model as qubits. 

To summarize, Shor's model views the near horizon region of the black hole as a computational network composed of many cells with order one qubits per cell and with inter-cell dynamics constrained by the causal structure of the black hole spacetime. Moreover, the model is such that the Bekenstein-Hawking entropy of the black hole roughly matches the total entropy in all the cells. We also permit unspecified high-energy ``quantum gravity'' dynamics within and between neighboring cells, but whatever this unknown dynamics is, it is constrained to obey the large-scale causal structure of the black hole spacetime.

\subsection{Notions of scrambling and a potential puzzle}

Starting from the above model of black hole dynamics, Shor then proceeds to bound the time-scales for various notions of information scrambling. Weak scrambling is roughly defined as the success condition of the Hayden-Preskill protocol~\cite{Hayden:2007black}. 
The timescale for weak scrambling is expected to be $\beta \log S \sim M \log M$ and can be measured using OTOCs~\cite{Yoshida:2017efficient}. 
This timescale is compatible with the cell picture above since a small number of qubits can be transported from any point on the horizon to any other point using a path with only $\sim \log M$ cells. This is done by first sending the qubits up to the photon sphere (where there are a small number of cells and information can move around the black hole in a few steps), and then back down towards the horizon. Since it only takes $\sim \log M$ steps to reach the photon sphere from any point in the network, it follows that any pair of points can be connected in Schwarzschild time of order $M \log M$ (the basic unit of time multiplied by the radial number of cells). We caution the reader that we are not claiming this is the right way to understand weak scrambling, only that the basic causal constraints of the spacetime do not immediately forbid this scenario. The cell picture also demonstrates how energy and charge can spread exponentially fast across the horizon as they fall in towards the black hole (something known long ago from the membrane paradigm point of view).

By contrast, strong scrambling refers to the situation in which parts of the quantum system, say, the upper and lower hemispheres of the black hole, are near maximally entangled~\cite{Lashkari:2011towards}. 
Nearly maximal means the entanglement is no more than a few bits away from its late time value (which is of order the entropy $S$).
Starting from a product state, Shor argues that if the causal structure of spacetime is given by the cell model, even at the stretched horizon, then there is no way that the strong scrambling time can be of order $M \log M$. 
This is because we need to move order $S\sim M^2$ qubits from top to bottom, but this takes a time of order $M^2$ when using the stretched horizon degrees of freedom. This is because we need to transport $M^2$ bits but we have only $M$ channels between the northern and southern hemispheres (the number of cells along the equator) each of which can move $1$ qubit per unit time $\beta=M$. Hence, using each channel $M$ times, we can move $M^2$ bits, but each use takes Schwarzschild time $M$, so the total Schwarzschild time is $M^2$. One might think of using the higher up cells again to move qubits faster, but because there are so few high cells, there is a bottleneck to transporting lots of entanglement this way. Shor's conclusion is that unless causality is strongly modified at the horizon, black holes take at least a time of order $M^2$ to strong scramble. He also suggests that there isn't much evidence that the strong scrambling time is of order $M \log M$ (unlike the evidence for the weak scrambling time), so perhaps the causal structure doesn't need to be modified.

\subsection{Sharp puzzle for AdS black holes}

For black holes described by AdS/CFT duality, we can sharpen this puzzle because we can setup a situation in which the initial black hole state is significantly under-entangled and compute the resulting entanglement dynamics. It is not possible to have a sensible black hole geometry without any entanglement, but the initial entanglement can be a fraction of its expected maximum (thermal) value. 
One way to achieve this is to consider black hole microstates involving an end-of-the-world (EOW) brane~\cite{Maldacena:2001eternal, Kourkoulou:2017pure}. 
These are pure states of a single CFT which have an exterior region with the usual black hole geometry and partial interior region terminated by the EOW brane. 
Because the Ryu-Takayanagi (RT) surface (or really the Hubeny-Rangamani-Takayanagi surface) can end on the EOW brane~\cite{Ryu:2006holographic, Hubeny:2007a}, the initial entanglement of sufficiently large subregions of the CFT can differ substantially from the expected late time value.

Several works have computed the entanglement entropy in these microstate geometries. 
Specializing to the case of one hemisphere of the CFT with the other, these works found that the entanglement saturated in a time of order $\beta$ (for AdS-sized black holes).\footnote{In CFT2 on a circle, the condition for the RT surface to end on the EOW brane of tension $T$ in AdS units is $\sinh \frac{r_H \Delta \theta}{2L} \geq \cosh \frac{t_0 r_H}{L^2} \sqrt{\frac{1+T}{1-T}}$ where $t_0$ is the Schwarzchild time. Converting to CFT units, the saturation time is of order $\beta$ for $\Delta \theta=\pi$ and low tension.} This is a rather short time which doesn't depend on the entropy at all. 
One could worry that this is an artifact of the geometrical calculation, but from the RT point of view it is not clear where the problem is. Alternatively, and perhaps more physically, it could be that the large $N$ part of the entanglement does saturate rapidly in a time of order $\beta$ while a smaller order $N^0$ amount of entanglement only exponentially approaches its late time value. 
In any event, whether the strong scrambling time is of order $\beta$ or $\beta \log S$, it is clearly much smaller than the analog of $M^2$ in the Schwarzchild case. 

If we hypothesize that this timescale also applies to Schwarzschild black holes, then based on the discussion of the cell model, the causal structure of spacetime near the black hole cannot be given by the usual Schwarzchild metric (or another assumption must fail). 
Our proposed resolution is hopefully intuitive at this point. 
We postulate that at the stretched horizon where the local temperature is Planck-scale, the underlying microscopic quantum gravity degrees of freedom have been exposed. 
These are represented by some chaotic non-locally interacting system which has no local structure. 
However, as we argued extensively above in a model known to contain a gravity sector, the non-locality at the horizon would not have to strongly modify the causal structure further away. 
In particular, we could consistently maintain both that the system scrambles rapidly and that light outside the horizon continues to move just as dictated by the Schwarzchild metric (provided $M$ is large).

Hence, we are proposing that the degrees of freedom at the black hole horizon are fundamentally non-local from the point of view of an exterior observer. 
Nevertheless, no significant violation of causality would be observed in the weakly coupled degrees of freedom outside the horizon provided the black hole is large. 
Our theory does predict that small black holes could exhibit significant violations of causality, and it would be interesting to attempt to devise an experiment to test this prediction.


We now discuss quantitatively the case of an AdS black hole. 
Here we consider four dimensions for concreteness, but the puzzle exists in any dimension. The AdS black hole metric is
\bea
ds^2=- \frac{r^2 - r_+^2}{l^2} dt^2 + \frac{l^2}{r^2 - r_+^2} dr^2 + r^2 d\theta^2 + r^2 \sin^2 \theta d\phi^2,
\eea
where $l$ is the AdS radius and $r_+$ is the horizon radius. It is not hard to see the corresponding temperature observed by an asymptotic observer and the entropy are given by
\bea
\beta = \frac{2\pi l^2}{r_+}, \quad S = \frac{4\pi r_+^2 }{4G} \equiv \frac{r_+^2}{l_\text{pl}^2},
\eea
with $l_\text{pl}$ denoting the Planck length. 

To model the scrambling dynamics outside the black hole, similar to what Shor has considered, we divide the space outside the horizon into cells whose light-cross time is $\beta$ which is the unique unit in the theory. 
From this definition, the width and height of a cell at radius $r$ are, respectively, 
\bea
	\Delta x = 2\pi l \sqrt{ \frac{r^2}{r_+^2}-1}, \quad \Delta r = \frac{2\pi (r^2 - r_+^2) }{r_+}.
\eea
Accordingly, the number of cells at radius $r$ is
\bea
	N(r) = \frac{4\pi r^2}{\Delta x^2} = \frac{r^2}{\pi l^2} \frac1{\frac{r^2}{r_+^2} - 1}.
\eea
The number of cells decreases as $r$ increases. At $r_\text{max}=\sqrt{\frac{\pi}{\pi - r_+^2/l^2}} r_+ $ there is only a cell left (suppose $r_+ < \sqrt\pi l$ as we are mainly interested in the scaling regime with $r_+/l \sim 1$). Thus, $r_\text{max}$ is the outermost layer that contains only a cell.

The number of layers from $r_0$ to $r$ is
\bea
	L(r_0,r) = \int_{r_0}^r \frac{dr}{\Delta r} = \frac1{4\pi} \log \left( \frac{r-r_+}{r+ r_+} \frac{r_0 + r_+}{r_0 - r_+} \right),
\eea
where $r_0 > r_+$ is the radius of the stretched horizon that is determined later.

Having in mind how these cells are distributed around the black hole, we ask how capable each cell is of processing information. 
We assume the information can be modeled by black-body radiations, and in three spatial dimensions the black-body radiations in a cell of volume $V$ contain entropy
\bea
	S = \frac{4\pi^2}{45} V T_\text{loc}^3,
\eea
where $T_\text{loc}$ is the local temperature inside the cell. The local temperature at $r$ is given by a blue shift from the temperature at infinity, namely,
\bea
	T_\text{loc}(r) = \frac{r_+}{2\pi l^2} \frac1{\sqrt{1-r_+^2/r^2}}.
\eea
So the entropy of each cell at radius $r$ is
\bea
	\Delta S = \frac{4\pi^2}{45} \Delta x^3 T_\text{loc}^3 = \frac{r^3}{l^3}.
\eea
Although the entropy grows with the radius, because all layers of cells are concentrated near the horizon with even the outermost layer located at $r_\text{max} \sim r_+ \sim l$, each cell can only processes $\O(1)$ information.

The final question is where the stretched horizon is? Since we assume all information of the black hole is carried by the black-body radiation outside the horizon, in order to make up the black hole total entropy $S = r_+^2/l_\text{pl}^2$, the stretched horizon is given by following equation
\bea
	N(r_0) \Delta S = \frac{r_+^2}{l_\text{pl}^2},
\eea
where the left-hand side is the total amount of information carried by the black-body radiation in cells, and the right-hand side is the total entropy of the black hole. To solve this equation, we assume $r_0 = r_+ + h$, $h\ll r_+$, then it leads to stretched horizon,
\bea
	h \approx \frac{2\pi}{45} \frac{l_\text{pl}^2 r_+^4}{l^5}.
\eea
Since $l_\text{pl} \ll l \approx r_+$, the assumption is justified. The total number of layers of cells ranging from the stretched horizon $r_0$ to the outermost cell $r_\text{max}$ is given by
\bea
	L(r_0, r_\text{max}) = \frac1{4\pi} \log \left( \frac{r_\text{max} - r_+}{r_\text{max} + r_+}\frac{r_0 + r_+}{r_0 - r_+} \right) \approx \log \frac{l^5}{l_\text{pl}^2 r_+^3},
\eea
where in the last step we ignored all inessential numerical factors.

Now we have all the ingredients needed to estimate the scrambling time. 
To transmit a few bits of information from, say, the northern pole of the stretched horizon to the southern pole, one way to do it is to send the information up to the outermost layer, and then it to the southern hemisphere at the outermost layer, and finally send it back down to stretch horizon. 
Since each cell can process information its information in time $\beta$, for $\O(1)$ qubits, this method costs time
\bea
	t_\text{weak}^* = \beta L(r_0, r_\text{max}) = \frac{l^2}{r_+} \log \frac{l^5}{l_\text{pl}^2 r_+^3} \approx l \log \frac{l^2}{l_\text{pl}^2},
\eea
where in the last step we used the fact that we are mainly interested in the region $r_+/l \sim 1$. 
Since we know $S \sim l^2/l_\text{pl}^2$, the scrambling time is of order $\beta \log S$, which is consistent with other definitions of the weak scrambling time, e.g., from OTOCs.

However, for strong scrambling time, which is the time to scramble a nearly unentangled state between two hemispheres into an almost maximally entangled state, the trajectory used for the weak scrambling estimate is insufficient. This is because the outer layers have limited capacity as quantum channels since they contain an order one number of cells each of which can process only an order one number of bits per thermal time. More explicitly, in order to scramble the state, one needs to send order $S \sim r_+^2 / l_\text{pl}^2$ qubits from the northern hemisphere to the southern hemisphere. Using the outermost layer, the scrambling time would be
\bea
	\frac{r_+^2}{ l_\text{pl}^2} \times t_\text{weak}^\ast \approx \frac{l^3}{l_\text{pl}^2} \log \frac{l^2}{l_\text{pl}^2}.
\eea

A more efficient way to move the qubits is directly through degrees of freedom at the stretched horizon. 
According to the distribution of the cells, the quantum channels provided by cells in the equator between the two hemispheres can process $r_+/l_\text{pl}$ qubits per time step. Thus, to transmit $S \sim r_+^2/l_\text{pl}^2$ information we need $S/(r_+/l_\text{pl}) \sim r_+/l_\text{pl}$ steps where each step takes time $\beta$. Using the stretched horizon degrees of freedom to scramble the information therefore take a time
\bea
	t_\text{str}^\ast \sim \beta \times \frac{r_+}{l_\text{pl}} =\frac{2\pi l^2}{l_\text{pl}}.
\eea
where the subscript indicates the strong version of the scrambling time. 
Note that this is much faster than going via the outermost layer, but still larger than the weak definition of scrambling time at large $l$.

So, we basically get the same puzzle in the Schwarzschild-AdS black hole as in the Schwarzschild black hole if the black hole is big enough, which is not a surprise because they have the same near horizon causal structure.  Now comes the crucial advantage of the AdS setup: we can compute the strong scrambling time using AdS/CFT. By preparing an under-entangled initial state in the CFT, one which is dual to a black hole with an ETW brane behind the horizon, the geometric calculation gives a CFT entropy which saturates after a time of order $\beta \sim l$~\cite{Hartman_2013,Cooper_2019}. Assuming the CFT entropy can be identified with the black hole entropy (including the thermal atmosphere) in the usual way, we have a sharp contradiction.

Hence, while it has long been known that the thermal atmosphere contains roughly enough entropy to account for the Bekenstein-Hawking entropy of the black hole, the analysis above shows that the causal structure of the thermal atmosphere cannot account for the strong scrambling time of the black hole. Of course, this conclusion relies on AdS/CFT and, in particular, on the identification of CFT entropy with black hole entropy along with the Ryu-Takayanagi formula. Given these assumptions, one cannot maintain that the information content of the black hole comes entirely from quantum fields outside the horizon that obey the causal structure. Of course, the AdS/CFT calculation of the entropy explicitly references the interior, so it is hard to directly compare to with the exterior-only view in Shor's puzzle. We propose that to have a consistent exterior-only picture of the black hole's information dynamics, there must be quantum gravity degrees of freedom on the stretched horizon that are inherently non-local. By appealing to chaos-protected locality, such non-locality can be perfectly consistent with locality for simple degrees of freedom, like those in the thermal atmosphere. 

\section{Discussion}

In this paper, we established the existence of a phenomenon that we dubbed chaos-protected locality in which strongly non-local interactions, if sufficiently chaotic, can leave other local structures approximately intact. We demonstrated the physics in a simple model built from the SYK model, but we expect that the lessons are more general. For example, an all-to-all Brownian circuit model weakly coupled to a system with simple propagating particle or wave excitations should also preserve locality for the simple degrees of freedom. We conjecture that the same is true for matrix theories, and it would be interesting to verify this.

It would also be interesting to further develop the bulk computational model with the non-local degrees of freedom explicitly included. There are many questions to explore. For example, what happens to these degrees of freedom away from the black hole? How do they interplay with Lorentz invariance and the special frame defined by the black hole? Can they be explicitly related to underlying matrix degrees of freedom in models of AdS/CFT?

We also know that for AdS black holes that are significantly larger than the AdS radius, the dynamics contains a mix of local and non-local elements, i.e. both Lyapunov growth of OTOCs as well as spatial spread at the butterfly speed. The non-local interactions we proposed here should be local beyond some length scale, and it would be interesting to give a formula for this length scale. We can identify it simple cases from our knowledge of the CFT structure, but a general bulk principle is also desirable.

\section*{Acknowledgements}

This work is supported by the Simons Foundation via the It From Qubit Collaboration. The work of
BGS is also supported in part by the AFOSR under grant number FA9550-19-1-0360.
 
\appendix

\section{Free Majorana model} \label{append:free}
Without coupling to the chaotic Majorana, the local Majorana has the following action
\bea
	I_L = \sum_{r=1}^N \int d\t (\frac12 \psi_r \partial_\tau \psi_r - \ii w \psi_r \psi_{r+1}) = \frac12 \sum_k \int d\tau \psi_{-k}(\partial_\t + \varepsilon_k)\psi_k, \quad \varepsilon_k = 2w \sin k,
\eea
where we have assumed a periodic boundary condition, $\psi_{r+N}=\psi_r$, and used the Fourier transform
\bea
	\psi_k = \frac1{\sqrt N} \sum_k \psi_r e^{\ii k r}, \quad k = -\pi + \frac{2\pi n}{N}, \quad n= 1,...,N.
\eea
The imaginary time propagator is
\bea
	G_{\psi,0}(\omega,k) = \frac1{-\ii \omega + \varepsilon_k}, \quad G_{\psi,0}(\tau,k ) = -\frac{e^{- \varepsilon_k \tau}}{e^{\beta \varepsilon_k} + 1}.
\eea
The imaginary time propagator in the real space is obtained by inverse Fourier transform,
\bea
	G_{\psi,0}(\tau, r) = \frac1N \sum_k G_{\psi,0}(\tau, k) e^{\ii k r} = \sum_{s=\pm 1}  \int_{-\Lambda}^\Lambda \frac{dk}{2\pi} \frac{-e^{- s v k \tau}}{e^{\beta s v k} + 1},
\eea
where in the second step we make a continuum limit with linear dispersion $\varepsilon_k \approx s v k, v = 2w $, and $s$ counts for left and right propagating modes. We also introduce a cutoff of the momentum. By carefully taking $\Lambda \rightarrow \infty$ limit, we have the following imaginary time propagator
\bea
	G_{\psi,0}(\tau, r) = \frac1{2\pi v} \sum_{s=\pm 1} \frac{\pi}{\beta \sin \frac{\pi(\tau - \ii s r/v)}{\beta}}.
\eea
This is consistent with the fact that a free Majorana fermion at one dimension has scaling dimension one half. For the sake of complete discussion, the greater (lesser) propagator is
\bea
	\langle \psi_{r_1} (t_1) \psi_{r_2} (t_2)\rangle = -\frac{\ii}{2\pi v} \sum_{s=\pm 1} \frac{\pi}{\beta \sinh \frac{\pi(t_{12} - \ii s r_{12}/v)}{\beta}}, \quad t_{12} \equiv t_1-t_2, \quad r_{12} \equiv r_1 - r_2,
\eea
which leads to the retarded Green's function~(\ref{eq:retard}).

In the free theory, the four-point function can be obtained by Wick theorem,
\bea
	&& F^{(0)}_{r,r'}(\t_1, \t_2, \t_3, \t_4) = G_{\psi,0}(\t_{12}, 0) G_{\psi,0}(\t_{34},0) \nn \\
	&& - G_{\psi,0}(\t_{13}, r-r') G_{\psi,0}(\t_{24}, r-r') + G_{\psi,0}(\t_{14},r-r') G_{\psi,0}(\t_{23},{r-r'}).
\eea
By plugging the imaginary time propagator into the four-point function and taking an analytic continuation to $\t_1 \rightarrow \beta + \ii t$, $\t_2 \rightarrow \beta/2 + \ii t$, $\t_3 \rightarrow 3\beta/2$, $\t_4 \rightarrow \beta/2$, we arrive at
\bea
	&& F^{(0)}_{r,r'}(\t_1, \t_2, \t_3, \t_4) \nn \\
	&& = \frac1{(v \beta)^2}  \left[1 - \sum_{s = \pm 1} \left( \frac1{\cosh^2 \frac{2\pi}\beta (t- s \frac{r-r'}v)} + \frac1{\cosh \frac{2\pi}\beta t} \frac1{\cosh \frac{2\pi}\beta (t - s \frac{r-r'}v)} \right) \right]. 
\eea

\section{Large $q$ effective action of the SYK model} \label{append:large-q}
The effective action of the chaotic Majorana model without the coupling to the local Majorana is the same as the SYK model,
\bea
	- \frac{I_C}M &=& \log \Pf(\partial - \Sigma_\chi) + \int d\t_1 d\t_2\Big[ - \frac12 G_\chi \Sigma_\chi + \frac{\J^2}{4q^2} (2G_\chi)^q \Big].
\eea
Using the large $q$ ansatz, $G_\chi(\t_1, \t_2) = \frac12 \sgn(\t_1-\t_2) (1 + \frac1q g(\t_1,\t_2))$, the action can be simplified to 
\bea
	- \frac{I_C}{M} &=& \int d\t_1 d\t_2 \left( - \frac1{4q^2}  \partial_{\t_1} [G_0 g] \partial_{\t_2} [G_0 g] + \frac{\J^2}{4q^2}  e^{g(\tau_1,\tau_2)} \right).
\eea
The equation of motion and the solution is given by
\bea
	\partial_\tau^2 g = 2\J^2 e^g,  \quad g= \log \frac{\alpha^2}{\J^2 \sin^2(\alpha |\tau_{12}| + \gamma)}, \quad \alpha = \J \sin \gamma, \quad \gamma = \frac\pi2 - \frac{\alpha\beta}2.
\eea
So the saddle-point solution from large $q$ approximation is $\bar G_\chi(\tau) = \frac12 \sgn(\t)  \left( \frac{1}{\sin(\alpha |\tau| + \gamma)} \right)^{2/q}$.

Having the saddle-point solution, one can further expand the large $q$ effective action in terms of the fluctuation, $G_\chi(\tau_1, \tau_2) = \frac12 \sgn(\t_1-\t_2) [1 + \frac1q (g(\t_1,\t_2)+ \delta g(\t_1, \t_2))] $. We are interested in chaotic properties which is the following propagator for bilocal field at long times and out-of-time order, 
\bea
	\langle G_\chi(\tau_1, \tau_2) G_\chi(\tau_3, \tau_4) \rangle =\bar G_\chi(\tau_1, \tau_2) \bar G_\chi(\tau_3, \tau_4) + \frac1{4q^2}  \langle \delta g(\t_1, \t_2) \delta g(\t_3, \t_4) \rangle,
\eea
where it is convenient to choose $\t_1 \rightarrow \beta + \ii t$, $\t_2 \rightarrow \beta/2 + \ii t$, $\t_3 \rightarrow 3\beta/2$, $\t_4 \rightarrow \beta/2$. The last term indicates it is the propagator of the fluctuation $\delta g(\tau_1, \tau_2)$. We in principle can obtain the imaginary time propagator and analytically continue to real times, but to make the calculation easier, we analytically continue the action, 
\bea
	-\frac{I_C}{M} &=& -\frac{1}{16q^2 }\int dt_1 dt_2 \left( \partial_{t_1} \delta g \partial_{t_2} \delta g + \frac{2\alpha^2}{\cosh^2 \alpha t_{12}}  \delta g^2 \right), \\
	&=& \frac{1}{16q^2 }\int d\bar t dt \delta g  \left( \frac14 \partial_{\bar t}^2  - \partial_t^2 -  \frac{2\alpha^2}{\cosh^2 \alpha t}  \right) \delta g , 
\eea
where in the second line we transform time coordinates by $\bar t= \frac{t_1 + t_2}2$, $t = t_1 - t_2$. The potential term is a function of time difference, so $H = -\partial_t^2 -  \frac{2\alpha^2}{\cosh^2 \alpha t}$ can be mapped to a one dimensional particle living in coordinate $t$. One can expand the fluctuation in the eigenbasis of this particle. This potential has only one bound state with negative energy, which corresponds to the chaotic mode,
\bea
	H \psi(t) = - \alpha^2 \psi(t), \quad \psi(t) = \sqrt{\frac{\alpha}{2}} \frac1{\cosh \alpha t}.
\eea
Thus in terms of the wavefunction, we can get the chaotic mode ,
\bea
	-\frac{I_C}{M} = \frac{1}{16q^2 }\int d\bar t \delta g(\bar t)  \left( \frac14 \partial_{\bar t}^2  -\alpha^2 \right) \delta g(\bar t), \\
	\langle \delta g(t_1, t_2) \delta g(t_3, t_4) \rangle = -  \frac{8q^2 \cosh 2\alpha (\bar t_{12} - \bar t_{34})}{\cosh \alpha t_{12} \cosh \alpha t_{34}}  
\eea
where $\delta g(t_1, t_2) = \delta g(\bar t ) \psi(t)$. In terms of the large $q$ fluctuations, the OTOC reads
\bea
	\langle G_\chi(\tau_1, \tau_2) G_\chi(\tau_3, \tau_4) \rangle = \bar G_\chi(\tau_1, \tau_2) \bar G_\chi(\tau_3, \tau_4) - \frac12  \frac{\cosh 2\alpha (\bar t_{12} - \bar t_{34})}{\cosh \alpha t_{12} \cosh \alpha t_{34}}.
\eea

\section{Low energy analysis: An irrelevant perturbation at low energies} \label{append:conformal}

In this section, we analyze the effect from the local Majorana to the chaotic Majorana at low energies, i.e., we will set $M=N$ and focus on the low energy limit $\J \beta \gg 1$. And in this section, we take $\beta = 2\pi$ for simplicity and restore the dimension in the final result.

As we shown before, at low energy limit, the chaotic Majorana is dominated by the Schwarzian action,
\bea
	\frac{I_C}{N} = -\frac{\alpha_S}{\J} \int d\t \Sch \left( \tan \frac{ g(\t)}2, \t \right) =  \frac{\alpha_S}{2\J} \int d\tau \left( \big(\frac{g''}{g'} \big)^2 - \big(g' \big)^2\right).
\eea
where $g(\tau)$ is the reparametrization field for times. For the coupling between local and chaotic Majorana, we can again use the large-$N$ property to get
\bea\label{eq:correction2}
	\langle e^{- I_{LC}} \rangle &\approx& e^{\frac{V^2}{2p^2} \int d\t_1 d\t_2  G_\chi(\tau_1, \tau_2)^p \sum_r \langle \psi_r(\tau_1) \psi_r(\tau_2) \rangle} \\
	&=& \exp\left( N \frac{V^2}{2p^2} \frac1{\pi v} \int d\t_1 d\t_2  \frac{\pi}{\beta \sin \frac{\pi \tau_{12}}\beta} G_\chi(\tau_1, \tau_2)^p   \right).
\eea
Considering the reparametrization fluctuation at low energies, 
\bea
	G_\chi(\tau_1, \tau_2) \rightarrow  b  \left( \frac{g'(\tau_1) g'(\tau_2) }{\big(2\sin \frac{g(\tau_1) - g(\tau_2)}2 \big)^{2}} \right)^{1/q},
\eea
the effective action with backreactions reads
\bea
	\frac{\tilde I_C}N = \frac{\alpha_S}{2\J} \int d\tau \left( \big(\frac{g''}{g'} \big)^2 - \big(g' \big)^2\right) - \frac{V^2}{2p^2} \frac{b^p}{\pi v} \int d\t_1 d\t_2  \frac{1}{ 2\sin \frac{|\tau_{12}|}2} \left( \frac{g'(\tau_1) g'(\tau_2) }{\big(2\sin \frac{g(\tau_1) - g(\tau_2)}2 \big)^{2}} \right)^{\frac{p}q}.
\eea
Note here we do not consider the reparametrization of times in the propagator from the local Majorana, this is because in the path integral measure, we already integrate out the local Majorana field in~(\ref{eq:correction2}). The field left unintegrated is that of chaotic Majorana, where the reparametrization dominates at the low energy.

One can define $e^{\phi(\tau)} = g'(\tau)$ so that the effective action becomes 
\bea
	\frac{\tilde I_C}N &=& \frac{\alpha_S}{2\J} \int d\tau \left(  \phi'(\tau)^2 - e^{2\phi(\tau)} \right) - \frac{V^2}{2p^2} \frac{b^p}{\pi v} \int d\t_1 d\t_2  \frac{1}{ 2\sin \frac{|\tau_{12}|}2}  \frac{ e^{\frac{p}q[\phi(\tau_1 ) + \phi(\tau_2)]} }{\big(2\sin \frac12 \int_{\tau_1}^{\tau_2} d\tau e^{\phi(\tau)} \big)^{\frac{2p}q}}.
\eea
From the second term, we can infer that the correction has the maximal effect when $\tau_1 = \tau_2$. It is seemingly divergent at $\tau_1 = \tau_2$. However, we know this short-range divergence is unphysical and can be resolved by noting that for Majorana operator $ \lim_{\eta \rightarrow 0 }\psi_r(\tau+ \eta) \psi_r(\tau) = \frac12 $ (similar for the chaotic Majorana). We can focus on this largest correction and simplify the local Majorana propagator by $\frac1{2\sin \frac{\pi \tau_{12}}\beta} \sim \frac{\delta(\tau_{12})}{J}$, and use the ultraviolet value to cutoff the divergence from chaotic Majorana. The effective action becomes
\bea
	\frac{\tilde I_C}N &=&  \int d\tau \left(  \frac{\alpha_S}{2\J} \Big(\phi'(\tau)^2 - e^{2\phi(\tau)}\Big)  -  \frac{V^2}{p^2v} \frac{b^p}{2\pi J} e^{\frac{2p}q \phi(\tau)} \right).
\eea
The action describes a particle moving in the exponential potential, 
\bea
	H = \frac{\alpha_S}{2\J} \Big( -\partial_{\phi}^2 + e^{2\phi(\tau)} \Big) + \frac{V^2}{p^2v} \frac{b^p}{2\pi J}  e^{\frac{2p}q \phi(\tau)}.
\eea
This is equivalent to Liouville quantum mechanics if there is no correction. We are not able to solve the full Hamiltonian, but we can treat the correction from local Majorana as a perturbation. Without the correction, the eigenstate with eigenenergy $E_k = \frac{\alpha_S}{2\J}  k^2 $ is~\cite{Bagrets:2017power}
\bea
	\psi_k(\phi) = \N_k K_{\ii k}(e^{\phi}), \quad \N_k = \frac{\sqrt 2}{\Gamma(\ii k)},
\eea
where $K_{\ii k}(x)$ is the modified Bessel function of the second kind, and $\N_k$ is the normalization factor. These eigenstates are like the plane wave with momentum $k$. Then the energy correction from first order perturbation theory is
\bea
	E^{(1)}_k =   \frac{V^2}{p^2v} \frac{b^p}{2\pi J} \int d\phi \psi_k^\ast(\phi)  e^{\frac{2p}q \phi} \psi_k(\phi) \approx   \frac{V^2}{p^2v} \frac{b^p}{2\pi J}  \frac{\Gamma(1- \frac{p}q)^4}{2^{\frac{2p}q}\Gamma\big(2(1-\frac{p}q)\big)} k^2,
\eea
where it seems to be an innocent perturbation. Nevertheless, it indicates an instability at $p=q$: when $p > q$ the perturbation is irrelevant, so we have a finite correction, when $ p \rightarrow q$ the coefficient brows up, meaning that the unperturbed theory is controlled. Originally, the scaling properties for the perturbation are set by 
\bea
	\frac{V^2}{p^2v} \frac{b^p}{2\pi J} \sim \frac{V^2}{v} \left( \frac1{\beta J} \right)^{2p/q+1}.
\eea
When $p>q/2$, the perturbation is irrelevant at low energies. 
However, in the regime dominated by reparametrization mode, the requirement for meanful perturbation theory is more stringent, i.e., $p>q$, as seen from above.

\bibliographystyle{jhep}
\bibliography{references}

\end{document}